\documentclass[reqno]{amsart}

\def\Bra{\Big\langle}
\def\Ket{\Big\rangle}
\def\cHHar{\cH_{Har}^{\phi_t}}
\def\cHcor{\cH_{cor}^{\phi_t}}
\def\cHcd{\cH_{cor,d}^{\phi_t}}
\def\cHcod{\cH_{cor,od}^{\phi_t}}
\def\cHcodp{\cH_{cor,od}^{\phi_t,++}}
\def\cHcodms{\cH_{cor,od}^{\phi_s,--}}
\def\cHcodps{\cH_{cor,od}^{\phi_s,++}}
\def\cHcodm{\cH_{cor,od}^{\phi_t,--}}

\def\cHmf{\cH_{mf}^{\phi_t}}

\def\Nb{N_b}
\def\Pb{P_b}
\def\Qb{Q_b}
\def\vac{\Omega}

\def\C{{\mathbb C}}

\def\R{{\mathbb R}}
\def\Z{{\mathbb Z}}

\def\uk{{\underline{k}}}

\def\frH{{\mathfrak H}}

\def\cE{{\mathcal E}}
\def\cF{{\mathcal F}}
\def\cH{{\mathcal H}}

\def\cL{{\mathcal L}}
\def\cM{{\mathcal M}}

\def\cS{{\mathcal S}}
\def\cU{{\mathcal U}}
\def\cV{{\mathcal V}} 
\def\cW{{\mathcal W}} 

\def\1{{\bf 1}}

\def\eqnn{\begin{eqnarray*}}
\def\eeqnn{\end{eqnarray*}}
\def\eqn{\begin{eqnarray}}
\def\eeqn{\end{eqnarray}}

\def\prf{\begin{proof}}
\def\endprf{\end{proof}}

\theoremstyle{plain}
\newtheorem{theorem}{Theorem}[section]

\newtheorem{proposition}[theorem]{Proposition}

\newtheorem{lemma}[theorem]{Lemma}

\newtheorem{remark}[theorem]{Remark}

\numberwithin{equation}{section}

\begin{document}

\title[Quantum tracer particle in a boson gas]
{Mean field dynamics of a quantum tracer particle interacting with a boson gas}

\author{Thomas Chen}
\address[T. Chen]{Department of Mathematics, University of Texas at Austin, Austin TX 78712, USA}
\email{tc@math.utexas.edu}
\author{Avy Soffer}
\address[A. Soffer]{Department of Mathematics, Rutgers University, Piscataway, NJ 08854}
\email{soffer@math.rutgers.edu}

\begin{abstract}
We consider the dynamics of a heavy quantum tracer particle coupled to a non-relativistic boson field in $\R^3$.
The pair interactions of the bosons are of mean-field type, with coupling strength proportional to $\frac1N$ where $N$ is the expected particle number. Assuming that the mass of the tracer particle is proportional to $N$, we derive generalized Hartree equations in the limit $N\rightarrow\infty$. Moreover, we prove the global well-posedness of the associated Cauchy problem for sufficiently weak interaction potentials.
\end{abstract}

\maketitle

\section{Introduction of the model}

We consider a heavy quantum mechanical tracer particle coupled to a field of identical scalar bosons with two-particle interactions. The Hilbert space for the quantum tracer particle (with position variable $X\in\R^3$) is given by $L^2(\R^3)$. 
The boson Fock space is given by
\eqn
	\cF=\C\oplus\bigoplus_{n\geq1}\cF_n 
\eeqn
where
\eqn 
	\cF_n := (L^2(\R^3))^{\otimes_{sym}n} 
\eeqn
is the $n$-particle Hilbert space.
We denote the Fock vacuum by $\vac\in\cF$,
and introduce creation- and annihilation operators satisfying the canonical commutation relations
\eqn 
	[a_x,a_y^+]=\delta(x-y) 
	\;\;\;,\;\;\;
	[a_x,a_y]= 0
	\;\;\;,\;\;\;
	[a_x^*,a_y^*]= 0 \,,
\eeqn 
where $a_x\vac=0$ for all $x\in\R^3$.
The Hilbert space of the coupled system is given by 
\eqn
    \frH = L^2(\R^3)\otimes\cF \,.
\eeqn
We will study the time evolution of this system for initial data $\Phi_0\in\cH$ with a large but finite expected particle number, $\Bra\Phi_0,\1\otimes\Nb\Phi_0\Ket=N$, where 
\eqn 
    \Nb := \int dx \, a_x^+  a_x \,
\eeqn
is the boson number operator.
We assume that the bosons interact with one another via a mean field interaction potential $\frac\lambda {2N} v$, where $\lambda>0$ is a coupling constant. Moreover, we assume that the mass of the heavy tracer particle is $N$. Accordingly, the Hamiltonian of the system is given by
\eqn\label{eq-cHN-def-1} 
	\cH_N &:=& -\frac1{2N}\Delta_X\otimes\1+\1\otimes T 
	+ \int dx \, w(X-x)\otimes a_x^+ a_x 
	\nonumber\\
	&&\hspace{2cm}+\1\otimes \frac\lambda{2N}\int dx dy \, a_x^+a_x v(x-y) a_y^+ a_y
\eeqn
where
\eqn 
	T := \frac12 \int dx \, a_x^+(-\Delta_x a_x)
\eeqn
is the kinetic energy operator for the boson field,
$w$ is the potential energy accounting for the coupling between the tracer particle and the bosons, and $\frac\lambda {2N} v$ is the potential accounting for pair interactions between bosons.

This system exhibits a close formal similarity to the translation-invariant model in non-relativistic Quantum Electrodynamics (QED) describing a quantum mechanical electron coupled to the quantized electromagnetic radiation field. The framework that we will use in our analysis is strongly inspired by \cite{fr-1,fr-2} and \cite{bcffs2,bcfs-2,ch-4,chfro,cfp-1,cfp-2}. 

We define the conserved total momentum operator 
\eqn 
	P_{tot} = i\nabla_X \otimes \1 + \1 \otimes \Pb
\eeqn
where 
\eqn
	\Pb := \int dx \, a_x^+ (i\nabla_x a_x) \,
\eeqn
is the momentum operator for the boson field.
The Hamiltonian is translation invariant, $[\cH_N,P_{tot}]=0$. 
Accordingly, we consider the decomposition of $\frH$ as a fiber integral 
\eqn 
	\frH = \int_{\R^3}^\oplus dk \, \frH_k 
\eeqn
with respect to $P_{tot}$ where
the fiber Hilbert spaces $\frH_k$ are isomorphic to $\cF$, and invariant under the dynamics generated by the Hamiltonian $\cH_N$. 

For each fixed $k\in\R^3$, we consider the value $Nk$ of the conserved total momentum $P_{tot}$.
The restriction of $\cH_N$ to $\frH_k$ is given by the fiber Hamiltonian
\eqn\label{eq-cHN-def-1}
	\cH_N(k) 
	&:=& \frac1{2N}(Nk-\Pb)^2 + T + \int dx \, w(x) a_x^+ a_x 
	\nonumber\\
	&&\hspace{2cm}
	+ \frac\lambda{2N}\int dx dy \, a_x^+a_x v(x-y) a_y^+ a_y
	\nonumber\\
	&=:&\frac1{2N}(Nk-\Pb)^2 + T + W_1(0)+W_2 \,,
\eeqn
where 
\eqn
    W_1(y) := \int dx \, w(x-y) a_x^+ a_x \,.
\eeqn
We also introduce  
\eqn
    W := W_1(0)+ W_2 \,
\eeqn
for notational convenience.
We note that here, $x$ (under a slight abuse of notation) stands for the relative coordinate $x-X$, with origin located at $X=0$.
For a more detailed introduction to the fiber decomposition with respect to the conserved total momentum, we refer to \cite{bcffs2}.

We will in the sequel identify $\frH$ with $L^2(\R^3,\cF)$, and omit the tensor products from the notation in \eqref{eq-cHN-def-1}.
The solution of the Schr\"odinger equation on $\frH$ has the following form.

\begin{proposition}
Given $u\in L^2(\R^3)$ and $\Psi_{k,0}^\cF\in\cF$, let
\eqn\label{eq-Phi-u0-init-1}
	\Phi_{u,0}(X) := \int dk \, \widehat u(k) e^{iX\cdot(Nk-\Pb)}\Psi_{k,0}^\cF \; \in \frH \,.
\eeqn
Then, 
\eqn 
	\Phi_u(t,X) := \int dk \, \widehat u(k) 
	e^{i X\cdot(Nk-\Pb)}\Psi_k^\cF(t)
\eeqn
solves 
\eqn\label{eq-Phiu-Schrod-1}
    i\partial_t\Phi_u=\cH_N\Phi_u
\eeqn 
on $\frH$ with initial data $\Phi_u(0,X)= \Phi_{u,0}(X)\in\frH$, iff
$\Psi^\cF_k(t)\in\cF$ solves  
\eqn\label{eq-fiber-dyn-1}
	i\partial_t\Psi_k^\cF(t) \, = \, \cH_N(k)\Psi_k^\cF(t)
\eeqn
on $\cF$ with initial data $\Psi_k^\cF(0)=\Psi_{k,0}^\cF\in\cF$.
\end{proposition}

\prf
We have
\eqn
    \lefteqn{
    \Big(i\partial_t+\frac1{2N}\Delta_X \Big)\Phi_u(t,X)
    }
    \nonumber\\
    &=&\int dk \, \widehat u(k) 
	e^{i X\cdot(Nk-\Pb)}\Big(-\frac1{2N}(Nk-\Pb)^2\Big)
	\nonumber\\
	&&\hspace{1cm}
	+\frac1{2N}(Nk-\Pb)^2+T+W\Big)\Psi_k^\cF(t)
	\\
	&=&\int dk \, \widehat u(k) 
	\Big(e^{-i X\cdot \Pb}(T+W)e^{i X\cdot \Pb}\Big)e^{i X\cdot(Nk-\Pb)}\Psi_k^\cF(t) \,.
	\nonumber
\eeqn
Clearly,
\eqn
    e^{-i X\cdot \Pb} T e^{i X\cdot \Pb} = T 
\eeqn
and 
\eqn
    e^{-i X\cdot \Pb} a_x e^{i X\cdot \Pb} = a_{x+X}
    \;\;\;,\;\;\;
    e^{-i X\cdot \Pb} a_x^+ e^{i X\cdot\Pb} = a_{x+X}^+ \;\;\;,
\eeqn
as one sees from
\eqn 
    e^{-i X\cdot \Pb} a_x^+ e^{i X\cdot\Pb} 
    &=& \int dk e^{-ikx}e^{-i X\cdot \Pb} a_k^+ e^{i X\cdot\Pb} 
    \nonumber\\
    &=&\int dk e^{-ikx}a_k^+ e^{-i X\cdot (\Pb+k)} e^{i X\cdot\Pb} 
    \nonumber\\
    &=&\int dk e^{-ik(x+X)}a_k^+  \,.
\eeqn
Therefore,
\eqn
    e^{-i X\cdot \Pb} W_1 e^{i X\cdot \Pb} 
    &=&
    \int w(x) a_{x+X}^+a_{x+X} dx
    \nonumber\\
    &=&
    \int w(x-X) a_{x}^+a_{x} dx 
    \; = \;  W_1(X) \,,
\eeqn
and
\eqn
    e^{-i X\cdot \Pb} W_2 e^{i X\cdot \Pb} 
    &=&
    \frac{\lambda}{2N} \int a_{x+X}^+a_{x+X} v(x-y) a_{y+X}^+a_{y+X} dx dy
    \nonumber\\
    &=&
    \frac{\lambda}{2N} \int a_{x}^+a_{x} v(x-X-(y-X)) a_{y}^+a_{y} dx dy
    \nonumber\\ 
    &=& W_2 \,.
\eeqn
We thus conclude that
\eqn 
    \Big(i\partial_t+\frac1{2N}\Delta_X \Big)\Phi_u(t,X)
    =
    ( \, T+W_1(X)+W_2 \, ) \, \Phi_u(t,X) \,,
\eeqn
as claimed in \eqref{eq-Phiu-Schrod-1}.
\endprf
  
The main results proven in this paper can be summarized as follows. 

\subsection{The fiber ground state on $\cF_N$ for large $N$}

The fiber Hamiltonian $\cH_N(k)$ commutes with the number operator $\Nb$.
In Section \ref{sec-cFN-gs-1}, we study its restriction to the $N$-particle Fock space  $\cF_N$,
and prove that the fiber ground state energy,
\eqn
       E_N(k) &:=& {\rm infspec}\Big( \, H_N(k)\Big|_{\cF_N} \, \Big)
       \nonumber\\ 
       &=&
       \inf_{\Psi_N\in\cF_N;\|\Psi\|_{\cF_N}=1}\Bra \, \Psi_N \,,\, H_N(k) \, \Psi_N \, \Ket \,.
\eeqn 
satisfies the asymptotics
\eqn
    \lim_{N\rightarrow\infty}\frac{E_N(k)}{N} =
    \frac{k^2}4 + \inf_{\|\phi\|_{L^2}=1}\cE_0[\phi]  
\eeqn 
where
\eqn\label{eq-HEn-def-0}
    \cE_0[\phi]&:=& 
	\frac12\int dx \, |\nabla\phi(x)|^2
	+ \int dx \, w(x) \, |\phi(x)|^2
	\nonumber\\
	&&\hspace{2cm}
	+ \frac\lambda2 \int dx \, dy \, |\phi(x)|^2 v(x-y) |\phi(y)|^2 
\eeqn
is the Hartree energy functional; see Proposition \ref{prop-EN-asympt-1}.

The problem considered here corresponds to a mean field limit;
see \cite{lenaro} and the references quoted therein.
For the much more difficult case of a dilute gas and the Gross-Pitaevskii limit, we refer to  \cite{lise,liseyn,lisesoyn}.

\subsection{Coherent states and mean field limit as $N\rightarrow\infty$}

In Section \ref{sec-coh-mean-lim-1}, we derive the mean field limit of \eqref{eq-fiber-dyn-1} in the following sense.
We define the Weyl operator
\eqn 	
    \cW[\sqrt N\phi] &:=&
	\exp\Big( \, \sqrt N \int dx \, \big( \, 
	\phi(x)a_x^+ 
	- \overline{\phi(x)}a_x\, \big)\, \Big) \,,
\eeqn
and consider the solution of the Schr\"odinger equation \eqref{eq-Phiu-Schrod-1} with initial data given by a coherent state of the form
\eqn
	\cW[\sqrt N\phi_0]\,\vac
	=
	\exp\Big(-\frac N2\|\phi_0\|_2^2\Big)
	\exp\Big( \, \sqrt N \, \int dx \,
	\phi_0(x)a_x^+ \, \Big) \vac \,,  
\eeqn
for $\phi_0\in H^1(\R^3)$.  
   
For an arbitrary but fixed value $k\in\R^3$ of the conserved momentum, we assume that for some $T>0$, $\phi_t\in L^\infty_{t}H_x^3([0,T)\times\R^3)$ is the solution of
\eqn\label{eq-phi-pde-0}
    i\partial_t\phi_t \, = \, -
    \Big(k-(\phi_t,i\nabla\phi_t)\Big) \, i\nabla\phi_t  
    -\frac12 \Delta\phi_t +w \phi_t 
    + \lambda (v*|\phi_t|^2)\phi_t \,,
\eeqn 
with initial data $\phi_0\in H_x^3(\R^3)$. We introduce the scalar
\eqn\label{eq-Sfact-def-0}
    S(t,t')  &:=& N\int_{t'}^t ds \Big( \, -\frac{1}{2} k^2 + \frac12 (\phi_s,i\nabla\phi_s)^2
    \nonumber\\
    &&\hspace{2cm}
    + \frac\lambda2\int |\phi_s(x)|^2 v(x-y) |\phi_s(y)|^2 dx dy
    \, \Big) \,,
\eeqn
and a self-adjoint Hamiltonian quadratic in creation- and annihilation operators, of the form
\eqn\label{eq-cHmf-def-0}
    \cHmf(k)&:=&\cHHar(k)+\cHcor
\eeqn
where
\eqn 
    \cHHar(k)&:=&-\Big(k-(\phi_t,i\nabla\phi_t)\Big)\cdot\Pb 
    + T + W_1(0)
    \nonumber\\
    && 
    + \lambda
    \int | \phi_t(x)|^2 v(x-y) a_y^+ a_y  
    dx dy\,
\eeqn
is a generalized Hartree Hamiltonian which commutes with the particle number operator $\Nb$, and where
\eqn
    \cHcor
    &:=&
    \frac12\Big(a^+(i\nabla\phi_t)+a(i\nabla\phi_t)\Big)^2
    \nonumber\\
    &&
    + \lambda
    \int v(x-y)  \phi_t(x)\overline{\phi_t(y)} \, a^+_x a_y \, dx dy  
    \nonumber\\
    &&   
    + \frac{\lambda}{2}
    \int v(x-y)    \Big( \phi_t(x) \phi_t(y) a^+_x a^+_y + 
    \overline{\phi_t(x)} \overline{\phi_t(y)}a_y a_x  \Big) dx dy \,,
\eeqn
includes correlations which do not preserve the particle number.
Then,
\eqn\label{eq-cV-def-0}
    i\partial \cV(t,s) = \cHmf(k) \cV(t,s)
    \;\;\;,\;\;\;
    \cV(s,s)=\1 \,,
\eeqn
determines  the unitary flow $\cV(t,s)$ generated by $\cHmf(k)$.

Our main result in section  \ref{sec-coh-mean-lim-1} states that the limit
\eqn\label{eq-meanfield-N-lim-0}
    \lim_{N\rightarrow\infty}
    \Big\| e^{-it\cH_N(k)} \cW[\sqrt N\phi_0]\,\vac 
    - e^{-iS(t,0)}\cW[\sqrt N\phi_t] \cV(t,0)\,\vac \Big\|_{\cF} = 0
\eeqn
holds, under the assumption $v\in C^2(\R^3)$; see Theorem \ref {thm-cohconv-1}.
The more technical parts of the proof are presented in Section \ref{sec-meanfield-proof-1}.
Therefore, the solution of \eqref{eq-Phiu-Schrod-1}, with initial data \eqref{eq-Phi-u0-init-1} characterized by a coherent state $\Psi_{k,0}^\cF= \cW[\sqrt N\phi_0]\,\vac$, is given by
\eqn 
    \Phi_u(t,X) = \int dk \, \widehat u(k) 
	e^{i X\cdot(Nk-\Pb)} e^{-iS(t,0)} \cW[\sqrt N\phi_t] \cV(t,0)\,\vac + o_N(1)
\eeqn
asymptotically, as $N\rightarrow\infty$.

The convergence in Fock space proven in our work is closely related to \cite{grmama,grma-1,grma-2}.  
In our construction, the generator $\cHcor$ of $\cV(t,s)$ includes both a diagonal and an off-diagonal part, 
while in \cite{grmama,grma-1,grma-2} the term analogous to $\cV(t,s)$ has a purely off-diagonal generator. In the case considered here, the choice \eqref{eq-cV-def-0} with \eqref{eq-cHmf-def-0} allows us to efficiently control the operator $\frac1N(Nk- \Pb)^2$ in $\cH_N$ which is not present in the above mentioned works.
The Hamiltonian $\cHcor$ comprises the $O_N(1)$ terms of the quasifree reduction of $\cH_N$, in the sense of \cite{BBCFS}.
After completing this work, we noticed that a construction was used in \cite{lenasc} that is in part similar.

There is a vast literature on the derivation of time-dependent Hartree or nonlinear Schr\"odinger equations from an interacting boson field with mean field or Gross-Pitaevskii scaling. The first rigorous results were obtained in the pioneering papers \cite{he} and (a few years later) \cite{sp,Spohn81}; more recently, the works \cite{esy1,esy2,esy3,esy4}, and subsequently \cite{klma} and \cite{rosc}, motivated much of the current activity in the research area; we refer to
\cite{CHPS-1,chpa2,CPBBGKY,xchho1,xchho2,frknpi,frknsc,frtsya,GreSohSta-2012,grmama,grma-1,grma-2, kiscst,lenaro, lenasc,pick2}.

The mean field limit of a {\em classical} tracer particle coupled to a nonlinear Hartree equation was derived in \cite{defrpipi-1}.

\subsection{Analysis of the mean field equations}

In sections \ref{ssec-NLH-gs-1} and \ref{ssec-NLH-timedep-1}, we analyze the mean field equation \eqref{eq-phi-pde-0}.
In Section \ref{ssec-NLH-gs-1}, we determine the ground state for a conserved total momentum $k\in\R^3$, under the assumption that the Hartree energy functional for $k=0$ admits a minimizer; see Proposition \ref{prop-Hartree-gs-k-1}.

In Section \ref{ssec-NLH-timedep-1}, we study dispersive solutions to \eqref{eq-phi-pde-0}. In particular, we show that \eqref{eq-phi-pde-0} is unitarily equivalent to the nonlinear Hartree equation 
\eqn\label{eq-psi-main-eq-0}
    i\partial_t \psi = -\frac12\Delta \psi + w_\psi \psi + \lambda (v*|\psi|^2) \psi  
    \;\;\;,\;\;\;\psi(t=0)=\psi_0\equiv\phi_0 \in H_x^1(\R^3)
\eeqn
where 
\eqn 
    w_\psi(t,x) := w\Big(x-X_\psi(t)\Big)
\eeqn   
and 
\eqn 
    X_\psi(t) := \int_0^t ds \, \Big( k - (\psi, i\nabla \psi)(s)\Big)
\eeqn
is the expected trajectory of the tracer particle. 
 
In Theorem \ref{thm-psi-gwp-1}, we prove the global well-posedness for \eqref{eq-psi-main-eq-0} in the space
\eqn
    \psi \in   L_t^\infty H_x^1(\R\times\R^3)\cap L_t^{\frac{10}{3}} W_x^{1,\frac{10}{3}}(\R\times\R^3) \,,
\eeqn 
under the assumption that $\|w\|_{W_x^{2,\frac32}}<\infty$, and that
$\|w\|_{W_x^{1,\frac32}}$, $\|\lambda v\|_{W_x^{1,\frac32}}$ are sufficiently small.
As a corollary, we obtain global well-posedness for \eqref{eq-phi-pde-0}, in the sense stated in Theorem \ref{thm-phi-gwp-1}; in particular, $\phi(t,x)=\psi(t,x+X_\psi(t))$.
We note that under less restrictive assumptions on  $w$ and $\lambda v$, the problem can be controlled with methods developed in \cite{beso}.
We also note that $X_\psi(t)$ can be written as $X_\psi(t) = k t - (\psi, x \psi)$, and that it satisfies the Ehrenfest dynamics
\eqn\label{eq-Xpsi-Ehrenfest-1}
    \partial_t^2 X_\psi(t) 
    &=& \Big(\psi \, , \, \nabla(w_\psi+\lambda v*|\psi|^2) \psi \Big) 
    \nonumber\\
    &=&  \int dx(\nabla w)(x-X_\psi(t)) |\psi(x)|^2 \,.
\eeqn 
The term involving $v$ is zero because $v$ is even, see Remark \ref{rem-Ehrenfest-1}, below.

In particular, we find $\partial_t^2 X_\psi(t)=0$ in the special case where $\phi_0=Q_k$ is the nonlinear ground state of \eqref{eq-phi-pde-0}, with $\|Q_k\|_{L_x^2}=1$. This is because we have $Q_k=e^{-i\frac k2 x}Q_0$ where $Q_0$ is the rotationally symmetric minimizer of the Hartree functional \eqref{eq-HEn-def-1}, with $\|Q_0\|_{L_x^2}=1$; see Proposition \ref{prop-Hartree-gs-k-1}. Due to rotational symmetry of $Q_0$, we find that $X_\psi(t)=\frac k2 t$, and that with $\psi(t,x)=Q_k(x-\frac k2t)$, the r.h.s of \eqref{eq-Xpsi-Ehrenfest-1} is zero, so that $\partial_t^2 X_\psi(t)=0$. 

Clearly, \eqref{eq-Xpsi-Ehrenfest-1} describes a {\em  classical} tracer particle moving along the trajectory $X_\psi(t)\in\R^3$, coupled to a boson field described by the Hartree equation \eqref{eq-psi-main-eq-0}. A model of a similar type has been analyzed in \cite{frzh-1}, for which the emergence of Hamiltonian friction was established in certain cases in \cite{frzh-2,frzhso-1}.

\newpage
 
\section{The fiber ground state on $\cF_N$ for large $N$}
\label{sec-cFN-gs-1}

The fiber Hamiltonian $\cH_N(k)$ commutes with the number operator $\Nb$, and its restriction to the $N$-particle Fock space  $\cF_N$ is given by
\eqn
    H_N(k) := \cH_N(k)\Big|_{\cF_N} \,.
\eeqn
In this section, we will determine the asymptotics of its ground state energy in the limit of large $N$.
We define 
\eqn
       E_N(k) &:=& {\rm infspec}\Big( \, H_N(k)\Big|_{\cF_N} \, \Big)
       \nonumber\\ 
       &=&
       \inf_{\Psi_N\in\cF_N;\|\Psi\|_{\cF_N}=1}\Bra \, \Psi_N \,,\, H_N(k) \, \Psi_N \, \Ket \,.
\eeqn 
For large $N$, the following asymptotics hold.

\begin{proposition}\label{prop-EN-asympt-1}
The ground state energy of the fiber Hamiltonian $H_N(k)$ satisfies
\eqn
    \lim_{N\rightarrow\infty}\frac{E_N(k)}{N} =
    \frac{k^2}4 + \inf_{\|\phi\|_{L^2}=1}\cE_0[\phi]  
\eeqn 
where
\eqn\label{eq-HEn-def-1}
    \cE_0[\phi]&:=& 
	\frac12\int dx \, |\nabla\phi(x)|^2
	+ \int dx \, w(x) \, |\phi(x)|^2
	\nonumber\\
	&&\hspace{2cm}
	+ \frac\lambda2 \int dx \, dy \, |\phi(x)|^2 v(x-y) |\phi(y)|^2 
\eeqn
is the Hartree energy functional.
\end{proposition}

\prf
Let 
\eqn 
    (\tau_{\frac k2}\Psi_N)(x_1,\dots,x_N) := \exp\Big(-\frac{ik}{2}\sum_{j=1}^N x_j\Big) \, \Psi_N(x_1,\dots,x_N) \,.
\eeqn
Then, the kinetic energy part in \eqref{eq-cHN-def-1} yields
\eqn
    \lefteqn{
    \Bra \, \tau_{\frac k2}\Psi_N \,,\, \Big(\frac1{2N}(Nk-\Pb)^2 + T\Big) \, \tau_{\frac k2}\Psi_N \, \Ket 
    }
    \nonumber\\
    &=&
    \int \Big(\frac{1}{2N}\Big(Nk-\frac{Nk}2-\sum_{j=1}^N k_j\Big)^2
    +\frac12\sum_{j=1}^N \Big(k_j+\frac k2\Big)^2\Big)
    |\widehat\Psi_N(\uk_N)|^2 d\uk_N
    \nonumber\\
    &=&
    \int \Big(\frac{Nk^2}4+\frac{1}{2N}\Big(\sum_{j=1}^N k_j\Big)^2+\frac12\sum_{j=1}^N k_j^2\Big)
    |\widehat\Psi_N(\uk_N)|^2 d\uk_N
    \nonumber\\
    &=&\frac{Nk^2}4 + 
    \Bra \,  \Psi_N \,,\, \Big( \, \frac1{2N}\Pb^2 + T \, \Big) \, \Psi_N \, \Ket
    \,,
\eeqn
where $\uk_N:=(k_1,\dots,k_N)$ and $d\uk_N:=dk_1\cdots dk_N$,
while the interaction part yields
\eqn 
    \Bra \, \tau_{\frac k2}\Psi_N \,,\, W \, \tau_{\frac k2}\Psi_N \, \Ket 
    =
    \Bra \,  \Psi_N \,,\, W \, \Psi_N \, \Ket \,.
\eeqn
Thus, we obtain the lower bound
\eqn 
    E_N(k)
    &=&
    \inf_{\Psi_N\in\cF_N;\|\Psi_N\|_{\cF_N}=1}\Bra \, \Psi_N \,,\, H_N(k) \, \Psi_N \, \Ket
    \nonumber\\
    &=&
    \inf_{\Psi_N\in\cF_N;\|\Psi_N\|_{\cF_N}=1}
    \Bra \, \tau_{\frac k2}\Psi_N \,,\, H_N(k) \, \tau_{\frac k2}\Psi_N \, \Ket
    \nonumber\\
    &=&\frac{Nk^2}4 + 
    \inf_{\Psi_N\in\cF_N;\|\Psi_N\|_{\cF_N}=1}
    \Bra \, \Psi_N \,,\,  \Big( \, \frac1{2N}\Pb^2 + H_N(0) \, \Big) \, \Psi_N \, \Ket
    \nonumber\\
    &\geq&\frac{Nk^2}4 + E_N(0) \,.
\eeqn
Next, we determine an upper bound.

Let
\eqn
    \Psi_{N,\phi} := \frac{1}{\sqrt{N!}}(a^+(\phi))^N\vac
\eeqn
where $\phi\in H^1(\R^3)$ and $\|\phi\|_{L^2}=1$.
We choose
\eqn\label{eq-Qk-def-1}
    \phi = e^{-i\frac k2 x} Q_0 
\eeqn
where $Q_0 $ is the minimizer of the Hartree functional \eqref{eq-HEn-def-1} 
with mass $\|Q_0 \|_{L^2}=1$. 
Then, we find that 
\eqn\label{eq-EN0-upperbd-1}
     \lefteqn{
     \Bra \, \Psi_{N,\phi} \,,\, H_N(0) \, \Psi_{N,\phi} \, \Ket
     }
     \nonumber\\
     &=&
     \frac{Nk^2}4 +  
     \Bra \, \Psi_{N,Q_0 } \,,\, \frac1{2N}\Pb^2  \, \Psi_{N,Q_0 } \, \Ket
     +  
     \Bra \, \Psi_{N,Q_0 } \,,\,  H_N(0) \, \Psi_{N,Q_0 } \, \Ket
     \nonumber\\
     &=&
     \frac{Nk^2}4 +  
     \Bra \, \Psi_{N,Q_0 } \,,\, \frac1{2N}\Pb^2  \, \Psi_{N,Q_0 } \, \Ket
     +
     N\cE_0[Q_0 ] \,.
\eeqn
The minimizer of $\cE_0[\phi]$ is rotation symmetric, therefore
\eqn\label{eq-phi-rot-zero-1}
    \int k \, |\widehat Q_0 (k) |^2 dk = 0 \,,
\eeqn
and we have
\eqn\label{eq-Pbexp-id-1}
    \lefteqn{
    \Bra \, \Psi_{N,Q_0 } \,,\, \frac1{2N}\Pb^2  \, \Psi_{N,Q_0 } \, \Ket
    }
    \nonumber\\
    &=&
    \int  \frac1{2N}\Big(\sum_{j=1}^N k_j\Big)^2 \,  |\widehat Q_0 (k_1) |^2\cdots |\widehat Q_0 (k_N) |^2 d\uk_N
    \nonumber\\
    &=&
    \int  \frac1{2N}\Big(\sum_{j=1}^N k_j^2 \, \Big) \,  
    |\widehat Q_0 (k_1) |^2\cdots |\widehat Q_0 (k_N) |^2 d\uk_N
    \nonumber\\
    &=&
    \frac1{2} \int  k^2 \,  |\widehat Q_0 (k) |^2 \, dk \, < \infty \,.
\eeqn
Passing to the third line, we used that all terms whose integrands are proportional to $k_j \cdot k_\ell$, with $j\neq\ell$, vanish, due to \eqref{eq-phi-rot-zero-1}.
Notably, the term \eqref{eq-Pbexp-id-1} is $O(\frac1N)$ smaller than the other two terms on the last line of \eqref{eq-EN0-upperbd-1},
and we conclude that
\eqn\label{eq-EN-mainbd-1}
    \frac{k^2}4 + \frac{E_N(0)}N 
    \; \leq \; 
    \frac{E_N(k)}{N} 
    \; \leq \;
    \frac{k^2}4 
    + \frac1{2N} \int  k^2 \,  |\widehat{Q_0 }(k) |^2 \, dk 
    + \cE_0[Q_0 ] \,.
\eeqn
But as was proven in \cite{lenaro},
\eqn
    \lim_{N\rightarrow\infty}\frac{E_N(0)}N = \cE_0[Q_0 ] \,.
\eeqn 
Hence, \eqref{eq-EN-mainbd-1} implies that 
\eqn
    \lim_{N\rightarrow\infty}\frac{E_N(k)}{N} = \frac{k^2}4  
    + \cE_0[Q_0 ] \,,
\eeqn
as claimed.
\endprf


\section{Coherent states and mean field limit as $N\rightarrow\infty$}
\label{sec-coh-mean-lim-1}
 
In this section, we derive the dynamical mean field limit on Fock space.
In order to render the exposition more readable, we are presenting the core of the proof here, but provide the more technical parts of the proof later, in Section \ref{sec-meanfield-proof-1}.
In this paper, we are neither attempting to optimize the bounds on the convergence rates, nor the requirements on the potentials $w$ and $v$.

Let $\phi\in L^2(\R^3)$. In this section, it will be convenient to use the notation $\phi_t(x)\equiv \phi(t,x)$.
We define the Weyl operator
\eqn 	
    \cW[\sqrt N\phi] &:=&
	\exp\Big( \, \sqrt N \int dx \, \big( \, 
	\phi(x)a_x^+ 
	- \overline{\phi(x)}a_x\, \big)\, \Big)
\eeqn
We consider the solution of the Schr\"odinger equation \eqref{eq-Phiu-Schrod-1} with initial data given by a coherent state of the form
\eqn
	\cW[\sqrt N\phi_0]\,\vac
	=
	\exp\Big(-\frac N2\|\phi_0\|_2^2\Big)
	\exp\Big( \, \sqrt N \, \int dx \,
	\phi_0(x)a_x^+ \, \Big) \vac \,,  
\eeqn
for $\phi_0\in H^1(\R^3)$.

Moreover, we define a time-dependent mean-field Hamiltonian which is self-adjoint and quadratic in creation- and annihilation operators, of the form
\eqn
    \cHmf(k)&:=&\cHHar(k)+\cHcor
\eeqn
where the "diagonal term"
\eqn\label{eq-cHHar-def-1}
    \cHHar(k)&:=&-\Big(k-(\phi_t,i\nabla\phi_t)\Big)\cdot\Pb 
    + T + W_1(0)
    \nonumber\\
    && 
    + \lambda
    \int | \phi_t(x)|^2 v(x-y) a_y^+ a_y  
    dx dy\,
\eeqn
is the Hartree Hamiltonian which commutes with the particle number operator $\Nb$, and where the "off-diagonal term"
\eqn\label{eq-cHcor-def-1}
    \cHcor
    &:=&
    \frac12\Big(a^+(i\nabla\phi_t)+a(i\nabla\phi_t)\Big)^2
    \nonumber\\
    &&
    + \lambda
    \int v(x-y)  \phi_t(x)\overline{\phi_t(y)} \, a^+_x a_y \, dx dy  
    \\
    &&   
    + \frac{\lambda}{2}
    \int v(x-y)    \Big( \phi_t(x) \phi_t(y) a^+_x a^+_y + 
    \overline{\phi_t(x)} \overline{\phi_t(y)}a_y a_x  \Big) dx dy \,,
\eeqn
includes correlations which do not preserve the particle number.
We obtain the unitary flow $\cV(t,s)$, 
\eqn\label{eq-cV-def-1}
    i\partial_t \cV(t,s) = \cHmf(k) \cV(t,s)
    \;\;\;,\;\;\;
    \cV(s,s)=\1 \,,
\eeqn
generated by $\cHmf(k)$.

\begin{theorem}\label{thm-cohconv-1}
Let $k\in\R^3$. We assume that $v\in C^2(\R^3)$, and that for some $T>0$, $\phi_t\in L^\infty_{t}H_x^3([0,T)\times\R^3)$ is the solution of
\eqn\label{eq-phi-pde-1}
    i\partial_t\phi_t \, = \, -
    \Big(k-(\phi_t,i\nabla\phi_t)\Big) \, i\nabla\phi_t  
    -\frac12 \Delta\phi_t +w \phi_t 
    + \lambda (v*|\phi_t|^2)\phi_t \,,
\eeqn 
with initial data $\phi_0\in H_x^3(\R^3)$. Let 
\eqn\label{eq-Sfact-def-1}
    S(t,t')  &:=& N\int_{t'}^t ds \Big( \, -\frac{1}{2} k^2 + \frac12 (\phi_s,i\nabla\phi_s)^2
    \nonumber\\
    &&\hspace{2cm}
    + \frac\lambda2\int |\phi_s(x)|^2 v(x-y) |\phi_s(y)|^2 dx dy
    \, \Big) \,.
\eeqn
Then, the limit
\eqn\label{eq-meanfield-N-lim-1}
    \lim_{N\rightarrow\infty}
    \Big\| e^{-it\cH_N(k)} \cW[\sqrt N\phi_0]\,\vac 
    - e^{-iS(t,0)}\cW[\sqrt N\phi_t] \cV(t,0)\,\vac \Big\|_{\cF} = 0
\eeqn
holds.
\end{theorem}

\prf
We have
\eqn\label{eq-cFdiff-bd-1}
    \Big\| e^{-it\cH_N(k)} \cW[\sqrt N\phi_0]\,\vac 
    - e^{-iS(t,0)}\cW[\sqrt N\phi_t] \cV(t,0) \,\vac \Big\|_{\cF}^2
    =
    2( \, 1- M(t) \, )
\eeqn
where
\eqn
    M(t) &:=& Re \Bra \, e^{-it\cH_N(k)} \cW[\sqrt N\phi_0]
    \,\vac \, , \, e^{-iS(t,0)}\cW[\sqrt N\phi_t] \cV(t,0) \,\vac \Ket 
    \nonumber\\
    &=& Re \Bra \, 
     \,\vac \, , \,\cW^*[\sqrt N\phi_0] 
    e^{it\cH_N(k)} e^{-iS(t,0)}\cW[\sqrt N\phi_t] \cV(t,0) \,\vac \Ket \,.
\eeqn
One can easily verify that given \eqref{eq-phi-pde-1}, we have
\eqn 
    i\partial_t \cW[\sqrt N \phi_t] \, = \, [ \, \cHHar (k) \, , \, \cW[\sqrt N \phi_t] \, ] \,.
\eeqn
%
We consider the unitary flow
\eqn
    \cU(t,s) := \cW^*[\sqrt N\phi_s] e^{i(t-s)\cH_N(k)-iS(t,s)} \cW[\sqrt N\phi_t]
\eeqn
and introduce the selfadjoint operator
\eqn\label{eq-cL-def-1}   
    \cL_N^{\phi_t}(k) 
    &:=&\cW^*[\sqrt N\phi_t]\Big(\cH_N(k)-\partial_t S(t,0)
    \Big)\cW[\sqrt N\phi_t]
    \nonumber\\
    && -\cW^*[\sqrt N\phi_t][\cHHar(k),\cW[\sqrt N\phi_t]]- \cHmf(k)
    \nonumber\\
    &=&\cW^*[\sqrt N\phi_t] \cH_N(k) \cW[\sqrt N\phi_t] - \partial_t S(t,0)
    \nonumber\\
    &&-\cW^*[\sqrt N\phi_t]\cHHar(k)\cW[\sqrt N\phi_t] -\cHcor(k)  \,.
\eeqn
Then, it is clear that
\eqn\label{eq-cL-def-1}
    i\partial_t \Big( \cU_N(t,0)  \cV(t,0) \,\vac \Big) 
    = - \, \cU_N(t,0)  \cL_N^{\phi_t}(k) \,   \cV(t,0) \,\vac \,.
\eeqn
A straightforward but somewhat lengthy calculation shows that  
\eqn\label{eq-cL-expl-1}
    \cL_N^{\phi_t}(k)&=&
    \frac1{2\sqrt N}\Big(\Pb\cdot\big( a^+(i\nabla\phi_t)  +  a(i\nabla\phi_t) \big)  
    +\big( a^+(i\nabla\phi_t) +  a(i\nabla\phi_t) \big) \cdot\Pb\Big)
    \nonumber\\
    &&+ 
    \frac1{2N}\Pb^2
    \nonumber\\
    &&  
    + \frac{\lambda}{\sqrt N}
    \int v(x-y)  a_x^+\Big(  \overline{\phi_t(y)} a_y  
    + \phi_t(y)a^+_y  \Big) a_x \, dx dy
    \nonumber\\
    &&  
    + \frac{\lambda}{2N}
    \int v(x-y)  \,  a^+_x a^+_y a_y a_x  \, dx dy \,.
\eeqn 
This follows from Lemma \ref{lm-cL-1}.

Evidently,
\eqn
    M(t) &=& Re \Bra \,\vac \, , \, \cU_N(t,0) \, \cV(t,0)\,\vac  \Ket
    \nonumber\\
    &=&M(0)+Re\int_0^t \, ds \, \partial_s M(s)
    \nonumber\\
    &=&1-Re \Big\{ \, i\int_0^t \, ds \,  \Bra \,\vac \, , \, \cU_N(w,0)   \cL_N^{\phi_s}(k) 
    \, \cV(s,0) \,\vac  \Ket \, \Big\}
    \,.
\eeqn
It follows from the unitarity of $\cU_N(t,0)$ that
\eqn
    \Big| \, \Bra \,\vac \, , \, \cU_N(t,0)  \cL_N^{\phi_t}(k) 
    \, \cV(t,0) \,\vac  \Ket  \, \Big|
    \leq 
    \Big\| \, \cL_N^{\phi_t}(k) 
    \, \cV(t,0) \,\vac \, \Big\|_{\cF} \,.
\eeqn
We prove in Lemma \ref{lm-cL-bd-1} that
\eqn\label{eq-cL-bd-1}
    \Big\| \, \cL_N^{\phi_t}(k) 
    \, \cV(t,0) \,\vac \, \Big\|_{\cF} \leq C_0\frac{e^{C_1 t}}{\sqrt N} \,,
\eeqn
for some constants $C_0$, $C_1$ depending on $\|v\|_{ C^2(\R^3)}$ and $\|\phi_t\|_{L^\infty_{t}H_x^3([0,T)\times\R^3)}$.
Hence, we find that
\eqn
    |M(t)-1| 
    \, \leq \,  
    C_0 \int_0^t \frac{e^{C_1 s}}{\sqrt N} ds
    \, < \, 
    \frac{C_0}{C_1}\frac{e^{C_1 t}}{\sqrt N} \,.
\eeqn
We therefore conclude that for any $t>0$,
the lhs of \eqref{eq-cFdiff-bd-1} converges to zero in the limit $N\rightarrow\infty$.
\endprf

For the convergence to the mean field dynamics in Theorem \eqref{thm-cohconv-1}, we required that the solution to the mean field equations obtained in \eqref{eq-phi-pde-1},
\eqn\label{eq-phi-pde-1-2}
    i\partial_t\phi_t \, = \, -
    \Big(k-(\phi_t,i\nabla\phi_t)\Big) \, i\nabla\phi_t  
    -\frac12 \Delta\phi_t +w \phi_t 
    + \lambda (v*|\phi_t|^2)\phi_t \,,
\eeqn 
has regularity $\phi\in L_t^\infty H_x^3([0,T)\times\R^3)$, for $T>0$ (possibly $T=\infty$).

In Section \ref{ssec-NLH-gs-1} and Section \ref{ssec-NLH-timedep-1}, 
we study solutions to \eqref{eq-phi-pde-1-2} under less strict assumptions on the regularity, $\phi\in L_t^\infty H_x^1$, in the following two cases. 
In Section \ref{ssec-NLH-gs-1},  we assume that the interaction potentials $w$ and $v$ are
such that the standard Hartree equation with external potential $w$ possesses a ground state, and we construct the ground state solution to \eqref{eq-phi-pde-1-2}.
In Section \ref{ssec-NLH-timedep-1}, we assume that $w$ and $v$ do not allow for
bound states.
In this situation, we are considering dispersive solutions of \eqref{eq-phi-pde-1-2}, and prove local and global well-posedness under the assumption that $\|w\|_{W^{2,3/2}}<\infty$, and that $\|w\|_{W^{1,3/2}}$, $\|\lambda v\|_{ W^{1,3/2}}$ are sufficiently small.

\begin{remark}
Global dispersive solutions to \eqref{eq-phi-pde-1-2} with regularity $\phi\in L_t^\infty H_x^3(\R\times\R^3)$ can be constructed along the same lines as in our analysis in section  \ref{ssec-NLH-timedep-1}, under the assumption that $\|w\|_{W^{4,3/2}}<\infty$, and that $\|w\|_{W^{3,3/2}}$, $\|\lambda v\|_{ W^{3,3/2}}$ are sufficiently small. Because the arguments are straightforward, we leave this part as an exercise.
\end{remark}

\begin{remark}
For comparison, we note that the model usually encountered in the literature, describing the boson gas without tracer particle, is given by
\eqn\label{eq-cHN-def-2-0}
    \widetilde\cH_N:=T + \int dx \, w(x) a_x^+ a_x  
	+ \frac\lambda{2N}\int dx dy \, a_x^+a_x v(x-y) a_y^+ a_y \,.
\eeqn
That is, no term $\frac1{2N}(Nk-\Pb)^2$ appears here, \cite{he,rosc,grmama}.

Our constructions in this section apply to this case, too, but simplify. 
The mean field equation for $\phi$ is the standard Hartree equation,
$$
    i\partial_t\phi=-\Delta\phi+w\phi+\lambda (v*|\phi|^2)\phi
$$ 
with $\phi(0)=\phi_0\in H_x^1$.

The first term on the right hand side of \eqref{eq-cHHar-def-1} and the first term on the right hand side of  \eqref{eq-cHcor-def-1} are then absent.
Moreover, the terms on the first two lines on the r.h.s. of \eqref{eq-cL-expl-1} are absent, yielding
\eqn\label{eq-cL-expl-2-0}
    \widetilde\cL_N^{\phi_t}&:=& \frac{\lambda}{\sqrt N}
    \int v(x-y)  a_x^+\Big(  \overline{\phi_t(y)} a_y  
    + \phi_t(y)a^+_y  \Big) a_x \, dx dy
    \nonumber\\
    &&  
    + \frac{\lambda}{2N}
    \int v(x-y)  \,  a^+_x a^+_y a_y a_x  \, dx dy \,.
\eeqn 
As a consequence, a bound of the form \eqref{eq-cL-bd-1},
\eqn\label{eq-cL-bd-2-0}
    \Big\| \,\widetilde \cL_N^{\phi_t}
    \, \cV(t,0) \,\vac \, \Big\|_{\cF} \leq C_0\frac{e^{C_1 t}}{\sqrt N} \,,
\eeqn
follows from Lemma \ref{lm-Nb-bd-1}, but Lemma \ref{lm-cL-bd-1} does not need to be invoked. In particular, the constants $C_0,C_1$ depend only on $\|\phi\|_{L_t^\infty H_x^1(I\times\R^3)}$; that is, only $H_x^1$-regularity of $\phi$ is required, not $H_x^3$-regularity. This implies a convergence result analogous to \eqref{eq-meanfield-N-lim-1}, but only requiring that $\|\phi\|_{L_t^\infty H_x^1(I\times\R^3)}<\infty$. Clearly, $I=\R$, if the Hartree equation for $\phi$ is globally well-posed.
\end{remark}


\section{Ground state for the generalized Hartree functional}
\label{ssec-NLH-gs-1}

We introduce the generalized Hartree energy functional corresponding to an arbitrary but fixed conserved momentum $k\in\R^3$,
\eqn\label{eq-cEk-def-1} 
	\cE_{k}[\phi] &:=& \frac1N \, \Bra \, \Phi_{N,\phi} \, , \, \cH_N(k)
	\, \Phi_{N,\phi} \, \Ket
	\nonumber\\
	&=& 
	\frac12\Big(k- \int dx \, \overline{\phi(x)}i\nabla_x \phi(x)\Big)^2
	+ \frac12\int dx \, |\nabla\phi(x)|^2
	\nonumber\\
	&&
	+ \int dx \, w(x) \, |\phi(x)|^2
	+ \frac\lambda2 \int dx \, dy \, |\phi(x)|^2 v(x-y) |\phi(y)|^2
	\nonumber\\
	&=&\frac12\Big(k- (\phi \,,\, i\nabla \phi)\Big)^2 + \cE_0[\phi] \,.
\eeqn
Here, $\cE_0[\phi]$ is the standard Hartree energy functional with external potential $w$.
We note that the scaling in the model is chosen in such a manner that $\cE_{k}[\phi]$ is independent of $N$.

Let $Q_0^{(\mu)}$ denote the minimizer of 
\eqn\label{eq-cSk-def-0}
    \cS_0[\phi]=\cE_0[\phi]-\mu\|\phi\|_{L^2}^2
\eeqn 
where $\mu$ is a Lagrange multiplier (the chemical potential) implementing the constraint that the $L^2$-mass $\|\phi\|_{L^2}^2=M$ is constant. 
Accordingly, $\mu$ depends on $M$, and
we denote by $\mu_0$ the value of $\mu$ for which $M=\|Q_0^{(\mu_0)}\|_{L^2}^2=1$.
For brevity, we write $Q_0 :=Q_0^{(\mu_0)}$, as we will only consider the case $M=1$.
By variation of \eqref{eq-cSk-def-0} in $\phi$, if follows that $Q_0 $ satisfies the stationary Hartree equation 
\eqn\label{eq-HartreeEq-1}
    \mu_0 Q_0  &=& 
    -\frac12\Delta Q_0  + wQ_0  +\lambda (v*|Q_0 |^2)Q_0 
\eeqn
where the value of $\mu_0$ is obtained from taking the inner product of \eqref{eq-HartreeEq-1} with $Q_0 $.

Likewise, for a nonzero total conserved momentum $k\in\R^3$, we consider the minimizer $Q_k $ of
\eqn\label{eq-cSk-def-1}
	\cS_k[\phi]:=\cE_k[\phi] - \mu \|\phi\|_{L^2}^2\,,
\eeqn 
under the constraint condition $\|Q_k \|_{L^2}^2=1$,
and we denote the corresponding value of the chemical potential by $\mu_k$.
By variation in $\phi$, it follows that the minimizer $Q_k $ satisfies 
\eqn\label{eq-Hartreek-def-1}
    \mu_k Q_k  =-\Big(k- (Q_k  \,,\, i\nabla Q_k )\Big)i\nabla Q_k 
    -\frac12\Delta Q_k  + wQ_k  +\lambda (v*|Q_k |^2)Q_k 
\eeqn
with $\|Q_k \|_{L^2}=1$. The value of $\mu_k$ is obtained from taking the inner product of \eqref{eq-Hartreek-def-1} with $Q_k $.  

\begin{proposition}\label{prop-Hartree-gs-k-1}
The vector  
\eqn\label{eq-phik-def-1}
    Q_k :=e^{-i\frac{k\cdot x}2}Q_0 
\eeqn
minimizes $\cS_0[\phi]=\cE_k[\phi]-\mu\|\phi\|_{L^2}^2$ with constraint $\|Q_k \|_{L^2}=1$, and satisfies \eqref{eq-Hartreek-def-1} with chemical potential
\eqn 
    \mu_k = \mu_0+\frac{k^2}{4}\,.
\eeqn
\end{proposition}

We remark that $\frac{k^2}4$ is the kinetic energy of a dressed particle, consisting of the tracer
particle together with a cloud of bosons, of total mass 2.

\prf
First of all, we verify from straightforward calculation that
\eqn 
    \cS_k[e^{-i\frac{k\cdot x}2}\phi] 
    &=& 
    \frac{k^2}{4} + \frac12 (\phi,i\nabla\phi)^2 + \cE_0[\phi] - \mu\|\phi\|_{L^2}^2
    \nonumber\\
    &\geq&
    \frac{k^2}{4}  + \cE_0[\phi] - \mu\|\phi\|_{L^2}^2 \,,
\eeqn
noting that for the choice of frequency $\frac k2$ in the exponent of \eqref{eq-phik-def-1},
terms linear in $i\nabla Q_0 $ on the right hand side cancel.
Therefore, 
\eqn 
    \inf_{\|\phi\|_{L^2}=1}
    \Big\{ \cS_k[e^{-i\frac{k\cdot x}2}\phi] \Big\}
    &\geq&
    \frac{k^2}{4}  + \inf_{\|\phi\|_{L^2}=1} \Big\{\cE_0[\phi] - \mu\|\phi\|_{L^2}^2\Big\}
    \nonumber\\
    &=& 
    \frac{k^2}{4} + \cE_0[Q_0 ] - \mu_0\|Q_0 \|_{L^2}^2
    \ \,.
\eeqn
On the other hand, because $Q_0 $ is spherically symmetric, $(Q_0 ,i\nabla Q_0 )=0$. Therefore,
\eqn 
    \cS_k[e^{-i\frac{k\cdot x}2}Q_0 ] 
    &=& 
    \frac{k^2}{4} + \frac12 (Q_0 ,i\nabla Q_0 )^2 + \cE_0[Q_0 ] - \mu\|Q_0 \|_{L^2}^2
    \nonumber\\
    &=&
    \frac{k^2}{4}  + \inf_{\|\phi\|_{L^2}=1} \Big\{\cE_0[\phi] - \mu\|\phi\|_{L^2}^2\Big\} 
\eeqn 
saturates the lower bound.

Furthermore, substituting $e^{-i\frac{k\cdot x}2}Q_0 $ for $Q_k $ in \eqref{eq-Hartreek-def-1}
yields
\eqn\label{eq-Hartreek-def-2}
    \mu_k Q_0  &=&\frac{k^2}{4}
    -\frac12 \Delta Q_0  + wQ_0  +\lambda (v*|Q_0 |^2)Q_0 
\eeqn
where we note that all terms proportional to $\nabla Q_0 $ cancel.
Comparing with \eqref{eq-HartreeEq-1}, we conclude that $\mu_k=\mu_0+\frac{k^2}{4}$, as claimed.
\endprf

\subsection{Mean field limit for the ground state}

Given $\phi_t=Q_k$ for some $k\in\R^3$, $\forall t\in\R$, \eqref{eq-phi-pde-1} 
is evidently solved by \eqref{eq-Hartreek-def-1}, and the expressions appearing in Theorem \ref{thm-cohconv-1} simplify as follows. 
The Hamiltonian
$\cHmf(k)=\cHHar(k)+\cHcor$
becomes time-independent, with
\eqn 
    \cHHar(k)&=& T + W_1(0) 
    + \lambda
    \int |  Q_0(x)|^2 v(x-y) a_y^+ a_y  
    dx dy\, ,
\eeqn
and
\eqn
    \cH_{cor}^{ Q_k}
    &=&
    \frac12\Big(a^+(i\nabla Q_k)+a(i\nabla Q_k)\Big)^2
    \nonumber\\
    &&
    + \lambda
    \int v(x-y)   Q_k(x)\overline{ Q_k(y)} \, a^+_x a_y \, dx dy  
    \nonumber\\
    &&   
    + \frac{\lambda}{2}
    \int v(x-y)    \Big(  Q_k(x)  Q_k(y) a^+_x a^+_y + 
    \overline{ Q_k(x)} \overline{ Q_k(y)}a_y a_x  \Big) dx dy \,.
    \nonumber 
\eeqn
Consequently, \eqref{eq-cV-def-1} simplifies to
\eqn
    \cV(t,s) = \exp\Big( - i(t-s)\cH_{mf}^{ Q_k}(k) \Big)\,,
\eeqn
and \eqref{eq-Sfact-def-1} simplifies to
\eqn 
    S(t,t') =  N (t-t') \Big( \,  \frac\lambda2\int | Q_0(x)|^2 v(x-y) | Q_0(y)|^2 dx dy
    \, \Big) \,,
\eeqn
inside the expression in \eqref{eq-meanfield-N-lim-1}. 
The nonlinear ground state $Q_0$ of the Hartree functional, normalized by $\| Q_0\|_{L_x^2}=1$ is, for $w$, $v\in C^1(\R^3)$, an element of $H_x^3(\R^3)$, see Lemma \ref{lm-Q0-H3-1}. Accordingly, $ Q_k\in H_x^3(\R^3)$, as required in  Theorem \ref{thm-cohconv-1}.

\begin{lemma}\label{lm-Q0-H3-1}
Assume that $w$, $v\in C^1(\R^3)$.
Let $Q_0$ denote the minimizer of the Hartree functional  $\cE_0[\,\cdot\,]$, satisfying \eqref {eq-HartreeEq-1} with $\mu_0<0$, and $\|Q_0\|_{L_x^2}=1$. Then, $ Q_0\in H_x^3(\R^3)$.
\end{lemma}

\prf
Given $\mu_0<0$ in \eqref{eq-HartreeEq-1}, we have the identity
\eqn 
    Q_0 = - (|\mu_0|-\Delta)^{-1} \Big(w Q_0 + \lambda (v*|Q_0|^2)Q_0\Big) \,
\eeqn
where $|\mu_0|-\Delta\geq|\mu_0|$ is strictly positive.
Therefore,
\eqn 
    \|Q_0\|_{H_x^3}&=&\Big\|(1-\Delta)^{\frac32}(|\mu_0|-\Delta)^{-1}
    \Big(w Q_0 + \lambda (v*|Q_0|^2)Q_0\Big)\Big\|_{L_x^2}
    \nonumber\\
    &\leq&
    |\mu_0|^{-1}
    \Big\| \, w Q_0 + \lambda (v*|Q_0|^2)Q_0 \, \Big\|_{H_x^1}
    \nonumber\\
    &\leq&|\mu_0|^{-1} (\|w\|_{C^1}+\|\lambda v\|_{C^1})\|Q_0\|_{H_x^1} \; < \; \infty \,,
\eeqn
using that $\|\lambda (v*|Q_0|^2)\|_{C^1}\leq \lambda \|v\|_{C^1}\|\,\|Q_0\|_{L_x^2}^2$, and $\|Q_0\|_{L_x^2}^2=1$.
\endprf


\section{Time-dependent mean field equations} 
\label{ssec-NLH-timedep-1}
 
The nonlinear dispersive PDE with conserved energy functional $\cE_k[\phi]$, for given $k\in\R^3$, is given by
\eqn\label{eq-phi-pde-2}
    i\partial_t\phi = -\Big(k-j_\phi(t)\Big) \, i\nabla\phi -\frac12 \Delta\phi +w \phi 
    + \lambda (v*|\phi|^2)\phi \,.
\eeqn
where 
\eqn 
    j_\phi(t):= (\phi(t),i\nabla\phi(t))
\eeqn
is the expected momentum (respectively, the current) determined by $\phi$.
The quantity $k-j_\phi$ is the momentum of the tracer particle, given that the 
bosons are in the coherent state parametrized by $\phi$, and the total conserved momentum is $k$.
We will prove the following global well-posedness result for the corresponding Cauchy problem.

\begin{theorem}\label{thm-phi-gwp-1}
Let $\phi_0\in H_x^1$. Then, there exists a unique global in time mild solution to \eqref{eq-phi-pde-2},
\eqn
    i\partial_t\phi = - \Big(k-j_\phi(t)\Big) \, i\nabla\phi -\frac12 \Delta\phi +w \phi 
    + \lambda (v*|\phi|^2)\phi \,,
\eeqn
with initial data $\phi(t=0)=\phi_0$, satisfying
\eqn 
    \|\phi\|_{L^\infty_t H_x^1(\R\times\R^3)} 
    + \|\tau_{X_\phi}\phi\|_{L_t^{\frac{10}{3}}W_x^{1,  \frac{10}{3}}(\R\times\R^3)}
    < \infty
\eeqn
where $(\tau_{X_\phi}\phi)(t,x):=\phi(x+X_\phi(t))$, and $X_\phi(t)=\int_0^t ds (k-(\phi,i\nabla\phi)(s))$.
In particular, $|\partial_t X_\phi(t)|<C\|\phi_0\|_{H_x^1}$, uniformly in $t\in\R$.
\end{theorem}

\prf
For any $\phi\in L^\infty_t H_x^1(\R\times\R^3)$,
\eqn 
    |j_\phi(t)| \leq \|\phi\|_{L^\infty_t H_x^1(\R\times\R^3)} < C 
\eeqn
is bounded, and therefore,  
\eqn 
    X_\phi(t) := \int_0^t ds(k-j_\phi(s))
\eeqn
is finite for every $t\in\R$. Consequently, $e^{iX_\phi(t) \cdot i\nabla}:H_x^1\rightarrow H_x^1$ is unitary for every $t\in\R$, and we may define
\eqn\label{eq-psi-def-1} 
    \psi(t,x) := e^{iX_\phi(t) \cdot i\nabla}\phi(t,x) \,.
\eeqn
The exponential generates translations in $x$-space by $X_\phi(t)$, yielding
\eqn\label{eq-psi-def-1-2} 
    \psi(t,x) =  \phi\Big(t,x-X_\phi(t) \Big) 
\eeqn 
as an equivalent expression.
Clearly, 
\eqn 
    j_\phi (t) = j_\psi(t) \,,
\eeqn
by unitarity of $e^{i\int_0^t ds(k-j_\phi(s))\cdot x}$ on $L^2(\R^3)$.
Therefore, 
\eqn\label{eq-Xphi-Xpsi-id-1}
    X_\phi(t) = X_\psi(t) \,,
\eeqn
and 
\eqn\label{eq-psi-Hartree-def-1}
    i\partial_t \psi = e^{iX_\phi(t) \cdot i\nabla}
    \Big( -\frac12 \Delta +w   
    + \lambda (v*|\phi|^2) \Big)
    e^{-iX_\phi(t) \cdot i\nabla} \psi \,,
\eeqn 
where we note that the first term on the r.h.s. of \eqref{eq-phi-pde-2} has been canceled by $(i\partial_tX_\phi(t))\phi$ obtained from the time derivative.
Noting that the operator $-\Delta$ is translation invariant, and
\eqn 
    \lefteqn{
    \Big(e^{iX_\phi(t) \cdot i\nabla} (v*|\phi|^2)  
    e^{-iX_\phi(t) \cdot i\nabla}\Big)(t,x)
    }
    \nonumber\\
    &=&
    \int v\Big(x-X_\phi(t)-y\Big) \, |\phi(t,y)|^2 \, dy
    \nonumber\\
    &=&
    \int v(x-y) \, |\phi(t,y-X_\phi(t))|^2 \, dy
    \nonumber\\
    &=&
    \big( v * |\psi|^2 \big)(t,x) \,,
\eeqn
we find that $\psi$ satisfies the nonlinear Hartree equation
\eqn\label{eq-psi-main-eq-1}
    i\partial_t \psi = -\frac12\Delta \psi + w_\psi \psi + \lambda (v*|\psi|^2) \psi  
    \;\;\;,\;\;\;\psi(t=0)=\psi_0\equiv\phi_0
\eeqn
where 
\eqn 
    w_\psi(t,x) := w\Big(x-X_\psi(t)\Big)
\eeqn    
is the potential $w$, translated by $X_\psi(t)$.

The claim of the theorem therefore follows from the global well-posedness of \eqref{eq-psi-main-eq-1} established in Theorem \ref{thm-psi-gwp-1}, below.
\endprf

\begin{remark}
The physical interpretation of \eqref{eq-psi-main-eq-1} is as follows. The field $\psi$ describes a self-interacting boson gas in mean field description which interacts with a point-like tracer particle traveling along a trajectory $X_\psi(t)$. The tracer particle creates an interaction potential $w$ which moves along  $X_\psi(t)$, here denoted by $w_\psi$. The momentum of the tracer particle, $\partial_t X_\psi(t)=k-(\psi,i\nabla\psi)$, together with the expected momentum of the boson field, $(\psi,i\nabla\psi)$, adds up to the conserved total momentum $k$. We note the close similarity to the equations studied in \cite{frzh-1} for a classical particle coupled to a boson gas.
\end{remark}
 
\begin{remark}
\label{rem-Ehrenfest-1}
The Ehrenfest dynamics of $X_\psi(t)$ is given by
\eqn\label{eq-Xpsi-Ehrenfest-2}
    \partial_t^2 X_\psi(t) = \Big(\psi \, , \, \nabla(w_\psi+\lambda v*|\psi|^2) \psi \Big)  \,,
\eeqn 
as stated in \eqref{eq-Xpsi-Ehrenfest-1}.
The term depending on $v$ is zero because, first of all,
\eqn\label{eq-vterm-1}
    \lefteqn{
    \Big(\psi \, , \, \nabla( v*|\psi|^2) \psi \Big) 
    }
    \nonumber\\
    &=& \int dx \, |\psi|^2 \nabla(v* |\psi|^2 )
    \nonumber\\
    &=& \int dx \, |\psi|^2 (v* \nabla |\psi|^2  ) \,.
\eeqn
On the other hand, using integration by parts,
\eqn\label{eq-vterm-2}
    \lefteqn{
    \int dx \, |\psi|^2 \, \nabla(v* |\psi|^2) 
    }
    \nonumber\\
    &=&
    - \int dx \, (\nabla |\psi|^2) \,  (v* |\psi|^2)
    \nonumber\\
    &=&
    - \int dx \, (v*\nabla|\psi|^2) \, |\psi|^2
\eeqn
where the last line holds because $v$ is a radial function, and thus even.
Comparing \eqref{eq-vterm-1} and \eqref{eq-vterm-2}, we find that 
$\int dx \, |\psi|^2 \nabla(v* |\psi|^2 )=-\int dx \, |\psi|^2 \nabla(v* |\psi|^2 )=0$.
Therefore, 
\eqn\label{eq-Xpsi-Ehrenfest-3}
    \partial_t^2 X_\psi(t) &=& \Big(\psi \, , \, \nabla w_\psi  \psi \Big) 
    \nonumber\\
    &=& \int dx \, (\nabla w)(x-X_\psi(t)) \, |\psi(x)|^2 \,,
\eeqn 
as claimed.
\end{remark}

A key advantage of \eqref{eq-psi-main-eq-1} over \eqref{eq-phi-pde-2} is that the Cauchy problem can largely be controlled with known dispersive tools for the analysis of the Hartree equation. A main difficulty is introduced by the dependence of the potential $w_\psi$ on $\psi$. Another complication arises from the fact that the energy is not conserved.

We construct local and global in time dispersive solutions under the assumption that $\|w\|_{W^{2,3/2}}<\infty$, and that $\|w\|_{W^{1,3/2}}$, $\|\lambda v\|_{ W^{1,3/2}}$ are sufficiently small. These conditions ensure, in accordance with the Birman-Schwinger principle, that neither $w$ nor $\lambda v$ create bound states. We require that  $w$ has one more derivative than $v$, to control the dependence of $w_\psi$ on $\psi$. 
First, we prove local well-posedness.  

\begin{theorem}\label{thm-lwp-main-1}
(Local well-posedness)
Let  $\psi_0\in H^1(\R^3)$ with $\|\psi_0\|_{L_x^2}=1$.
Assume that $\|w\|_{W_x^{2,\frac32}}<\infty$, and that
\eqn 
    \|w\|_{W_x^{1,\frac32}} + 3 \|\lambda v\|_{W_x^{1,\frac32}} <1 \,.
\eeqn
Then,  there exists a unique mild solution 
\eqn
    \psi \in   L_t^\infty H_x^1([0,T]\times\R^3)\cap L_t^{\frac{10}{3}} W_x^{1,\frac{10}{3}}([0,T]\times\R^3)
\eeqn 
to \eqref{eq-psi-Hartree-def-1} with initial condition $\psi(t=0)=\psi_0$, provided that  $T>0$ is sufficiently small.
\end{theorem}

\prf
We consider the map
\eqn 
    \cM \, : \, \psi \mapsto e^{it\Delta}\psi_0 + i\int_0^t ds \, e^{i(t-s)\Delta} 
    \Big((w_\psi \psi)(s)+\lambda ((v*|\psi|^2)\psi) (s)\Big) \,,
\eeqn
where we may assume that $\|\psi\|_{L_x^2}=1$.
Clearly, using the Strichartz and H\"older inequalities as in
\eqn 
    \| \int_0^t ds \, e^{i(t-s)\Delta} ( w_\psi \psi )(s) \|_{L^q_t L^r_x} &\leq&
    \|w_\psi \psi \|_{L_t^{\tilde q'}L_x^{\tilde r'}}
    \nonumber\\
    &\leq&\|w_\psi \psi \|_{L_t^{\frac{10}{3}}L_x^{\frac{30}{29}}}
    \nonumber\\
    &\leq& \|w_\psi\|_{L^\infty_t L_x^{\frac32}} \|\psi\|_{L_{t,x}^{\frac{10}{3}}} \,,
\eeqn
with $(q,r)$ and $(\tilde q, \tilde r)$ denoting arbitrary Strichartz admissible pairs, we find that, under inclusion of a derivative,
\eqn 
    \lefteqn{
    \|\cM[\psi]\|_{L_t^q W_x^{1,r}}
    }
    \\
    &\leq&\|\psi_0\|_{H^1} 
    + \|w_\psi\|_{L^\infty_t W^{1,\frac32}}\|\psi\|_{L_t^{\frac{10}{3}}W_x^{1,  \frac{10}{3}}}
    + \|\lambda v*|\psi|^2\|_{L^\infty_t W^{1,\frac32}}\|\psi\|_{L_t^{\frac{10}{3}}W_x^{1,  \frac{10}{3}}} \,.
    \nonumber
\eeqn
We use Young's inequality for convolutions in
\eqn 
    \|\lambda v*|\psi|^2 \|_{L^\infty_t W^{1,\frac32}} \leq \|\lambda v\|_{W^{1,\frac32}} \|\psi\|_{L^2}^2 \,,
\eeqn
and observe that
\eqn 
    \|w_\psi\|_{L^\infty_t W^{1,\frac32}} \leq \sup_{X} \|w(\,\bullet\,-X)\|_{W^{1,\frac32}} 
    = \|w\|_{W^{1,\frac32}} \,.
\eeqn
Therefore, 
\eqn\label{eq-cM-Strich-bd-1}
    \|\cM[\psi]\|_{L_t^q W_x^{1,r}} \leq \|\psi_0\|_{H^1}
    + \Big(\|w\|_{W^{1,\frac32}}+\|\lambda v\|_{W^{1,\frac32}}\Big)
    \|\psi\|_{L_t^{\frac{10}{3}}W_x^{1,  \frac{10}{3}}} \,,
\eeqn
for any Strichartz admissible pair $(q,r)$.
Consequently, writing $I:=[0,T]$, and defining the Banach space 
\eqn\label{eq-Yspace-def-1}
    Y(I) := L_t^\infty H_x^1(I\times\R^3)\cap L_t^{\frac{10}{3}} W_x^{1,\frac{10}{3}}(I\times\R^3)
\eeqn 
endowed with the norm
\eqn 
    \|f\|_{Y(I)} := \|f\|_{L_t^\infty H_x^1(I\times\R^3)} 
    + \|f\|_{L_t^{\frac{10}{3}} W_x^{1,  \frac{10}{3}}(I\times\R^3)} \,,
\eeqn
we find
\eqn\label{eq-cM-apriori-bd-1} 
    \|\cM[\psi]\|_{Y(I)} \leq 2\|\psi_0\|_{H_x^1}
    + \Big(\|w\|_{W_x^{1,\frac32}}+\|\lambda v\|_{W_x^{1,\frac32}}\Big) \|\psi\|_{Y(I)} \,.
\eeqn
Assuming that $\|w\|_{W_x^{1,\frac32}}+\|\lambda v\|_{W_x^{1,\frac32}}<1-\delta$, for some $\delta\in(0,1)$,
and defining $R:=2\delta^{-1}\|\psi_0\|_{H^1}$, we find that 
\eqn 
    \|\cM[\psi]\|_{Y(I)} \leq \delta R
    + (1-\delta) \|\psi\|_{Y(I)} \,.
\eeqn
Hence, the image of the ball $B_R(0)\subset Y$ under the map $\cM$ is contained in itself.

Next, we prove the contractivity of $\cM$. Given $\psi_1,\psi_2\in B_R(0)\subset Y$, we have
\eqn\label{eq-cM-contr-1}
    \|\cM[\psi_1]-\cM[\psi_2]\|_{Y(I)} 
    &\leq& \|w_{\psi_1}-w_{\psi_2}\|_{L^\infty_t W_x^{1,\frac32}}\|\psi_1\|_{Y(I)}
    \nonumber\\
    &&
    + \|w_{\psi_2}\|_{L^\infty_t W_x^{1,\frac32}}\|\psi_1-\psi_2\|_{Y(I)}
    \nonumber\\
    &&
    + \|\lambda v*(|\psi_1|^2-|\psi_2|^2)\|_{L^\infty_t W_x^{1,\frac32}}\|\psi_1\|_{Y(I)}
    \nonumber\\
    &&
    + \|\lambda v*|\psi_2|^2\|_{L^\infty_t W_x^{1,\frac32}}\|\psi_1-\psi_2\|_{Y(I)} \,.
\eeqn
To control the first term on the r.h.s. of \eqref{eq-cM-contr-1}, we have
\eqn 
    \|w_{\psi_1}-w_{\psi_2}\|_{L^\infty_t W_x^{1,\frac32}}
    &\leq&\| w\|_{W_x^{2,\frac32}}\sup_t|X_{\psi_1}(t)-X_{\psi_2}(t)|\,,
\eeqn
where
\eqn 
    |X_{\psi_1}(t)-X_{\psi_2}(t)| &\leq& \int_0^t ds \, |(\psi_1,i\nabla\psi_1)-(\psi_2,i\nabla\psi_2)|
    \nonumber\\
    &\leq&\int_0^t ds \, |(i\nabla(\psi_1-\psi_2),\psi_1)+(\psi_2,i\nabla(\psi_1-\psi_2))|
    \nonumber\\
    &\leq&t \, \|\psi_1-\psi_2\|_{L^\infty_t H_x^1}
    \Big(\|\psi_1\|_{L^\infty_t  L_x^2}+\|\psi_2\|_{L^\infty_t  L_x^2}\Big)
    \nonumber\\
    &\leq& 2 T \, \|\psi_1-\psi_2\|_{Y(I)}  
\eeqn
for $t\in I=[0,T]$.

To control the term on the third line on the r.h.s. of \eqref{eq-cM-contr-1}, we use
\eqn 
    \|\lambda v*(|\psi_1|^2-|\psi_2|^2)\|_{L^\infty_t W_x^{1,\frac32}}
    &\leq& 
    \|\lambda v\|_{W_x^{1,\frac32}} \||\psi_1|^2-|\psi_2|^2\|_{L^\infty_t L_x^1}
    \nonumber\\
    &\leq&
    \|\lambda v\|_{W_x^{1,\frac32}} \||\psi_1|+|\psi_2|\|_{L^\infty_t L_x^2}\|\psi_1-\psi_2\|_{L^\infty_t L_x^2}
    \nonumber\\
    &\leq&2 \|\lambda v\|_{W_x^{1,\frac32}} \|\psi_1-\psi_2\|_{Y(I)}
\eeqn
where $\||\psi_1|+|\psi_2|\|_{L_x^2}\leq\|\psi_1\|_{L_x^2}+\|\psi_2\|_{L_x^2}=2$.

Summarizing, we have 
\eqn\label{eq-cM-contr-2}
    \lefteqn{
    \|\cM[\psi_1]-\cM[\psi_2]\|_{Y(I)} 
    }
    \nonumber\\
    &\leq& 2 T \,\| w\|_{W_x^{2,\frac32}} \, 
    \|\psi_1\|_{Y(I)} \|\psi_1-\psi_2\|_{Y(I)} 
    \nonumber\\
    &&
    + \|w\|_{W_x^{1,\frac32}}\|\psi_1-\psi_2\|_{Y(I)}
    \nonumber\\
    &&
    + 2 \|\lambda v\|_{W_x^{1,\frac32}} \|\psi_1-\psi_2\|_{Y(I)}  
    \nonumber\\
    &&
    + \|\lambda v \|_{W_x^{1,\frac32}}\|\psi_1-\psi_2\|_{Y(I)} 
    \nonumber\\
    &\leq&
    \Big(2T R \| w\|_{W_x^{2,\frac32}} + \|w\|_{W_x^{1,\frac32}} + 3 \|\lambda v\|_{W_x^{1,\frac32}} \Big)
    \|\psi_1-\psi_2\|_{Y(I)} \,.
\eeqn
Therefore, $\cM$ is contractive on the ball $B_R(0)\subset Y$ if 
\eqn 
    2T R \| w\|_{W_x^{2,\frac32}} + \|w\|_{W_x^{1,\frac32}} + 3 \|\lambda v\|_{W_x^{1,\frac32}} <1 \,.
\eeqn
To this end, we require that 
\eqn 
    \|w\|_{W_x^{1,\frac32}} + 3 \|\lambda v\|_{W_x^{1,\frac32}} <1 \,,
\eeqn 
and that $T>0$ is sufficiently small (depending on $R$).
\endprf

We remark that the only place in the proof that requires a finite time $T>0$ is the control of $w_\psi$. 
The Strichartz estimates employed here remain valid with $\R$ instead of $I$.
We may therefore patch together local in time solutions using a global Strichartz inequality. 

\begin{theorem}\label{thm-psi-gwp-1}
(Global well-posedness)
Let  $\psi_0\in H^1(\R^3)$ with $\|\psi_0\|_{L_x^2}=1$.
Assume that $\|w\|_{W_x^{2,\frac32}}<\infty$, and that
\eqn 
    \|w\|_{W_x^{1,\frac32}} + 3 \|\lambda v\|_{W_x^{1,\frac32}} <1 \,.
\eeqn
Then, there exists a unique global mild solution $\psi\in Y(\R)$ to \eqref{eq-psi-Hartree-def-1} with initial condition $\psi(t=0)=\psi_0$.
In particular, it satisfies
\eqn\label{eq-psisol-unif-bd-1}
    \|\psi\|_{Y(\R)} \leq 
    2\Big(1- \|w\|_{W_x^{1,\frac32}} -  \|\lambda v\|_{W_x^{1,\frac32}}\Big)^{-1}\|\psi_0\|_{H_x^1} \,.
\eeqn
Moreover, 
\eqn\label{eq-parttXpsi-bd-1}
    |\partial_t X_\psi(t)|  
    &\leq& |k| + \Big(1- \|w\|_{W_x^{1,\frac32}} -  
    \|\lambda v\|_{W_x^{1,\frac32}}\Big)^{-1}\|\psi_0\|_{H_x^1} 
    \;\;,\;t\in\R \,;
\eeqn
that is, the momentum of the tracer particle is uniformly bounded in time.
\end{theorem}

\prf
The Strichartz estimate obtained in \ref{eq-cM-Strich-bd-1} holds globally in time. With $(q,r)=(\frac{10}3,\frac{10}3)$, it implies
\eqn\label{eq-cM-Strich-bd-2}
    \|\psi\|_{L_t^{\frac{10}{3}}W_x^{1,  \frac{10}{3}}(\R\times\R^3)} \leq \|\psi_0\|_{H^1}
    + \Big(\|w\|_{W^{1,\frac32}}+\|\lambda v\|_{W^{1,\frac32}}\Big)
    \|\psi\|_{L_t^{\frac{10}{3}}W_x^{1,  \frac{10}{3}}(\R\times\R^3)} \,,
\eeqn
respectively,
\eqn\label{eq-psi-Strich-bd-1}
    \|\psi\|_{L_t^{\frac{10}{3}}W_x^{1,  \frac{10}{3}}(\R\times\R^3)}\leq 
    \Big(1- \|w\|_{W_x^{1,\frac32}} -  \|\lambda v\|_{W_x^{1,\frac32}}\Big)^{-1}\|\psi_0\|_{H_x^1} \,.
\eeqn
We use this a priori bound to control the $L^\infty H_x^1$ norm of $\psi$.

Let $I_j:=[(j-1)T, jT]$.
The estimate \eqref{eq-cM-Strich-bd-1} with $(q,r)=(\infty,2)$, combined with \eqref{eq-psi-Strich-bd-1}, implies that  
\eqn\label{eq-psisol-unif-bd-1}
    \|\psi\|_{L_t^\infty H_x^1(I_1\times\R^3)} \leq 
    2\Big(1- \|w\|_{W_x^{1,\frac32}} -  \|\lambda v\|_{W_x^{1,\frac32}}\Big)^{-1}\|\psi_0\|_{H_x^1}\,, 
\eeqn
and hence,
\eqn\label{eq-psisol-unif-bd-2}
    \|\psi(t=2T)\|_{H_x^1(I_1\times\R^3)}\leq 
    \Big(1- \|w\|_{W_x^{1,\frac32}} -  \|\lambda v\|_{W_x^{1,\frac32}}\Big)^{-1}\|\psi_0\|_{H_x^1} \,.
\eeqn
Applying Theorem \ref{thm-lwp-main-1} for $I_2$ with initial data $\psi(t=2T)$ yields local well-posedness on $Y(I_2)$ with the same upper bound on $\|\psi(t=3T)\|_{H_x^1(I_1\times\R^3)}$ as in \eqref{eq-psisol-unif-bd-2}.

Iterating this argument for $I_j$, $j\in \Z$, we find that
\eqn\label{eq-Linfty-H1-psi-1}
    \|\psi\|_{L_t^\infty H_x^1(\R\times\R^3)}\leq 
    \Big(1- \|w\|_{W_x^{1,\frac32}} -  \|\lambda v\|_{W_x^{1,\frac32}}\Big)^{-1}\|\psi_0\|_{H_x^1} \,,
\eeqn
globally in time.

Therefore, we obtain  
\eqn\label{eq-cM-apriori-bd-2} 
    \|\psi\|_{Y(\R)} &=& 
    \|\psi\|_{L_t^\infty H_x^1(\R\times\R^3)}+\|\psi\|_{L_t^{\frac{10}{3}}W_x^{1,  \frac{10}{3}}(\R\times\R^3)}
    \nonumber\\
    &\leq& 
    2\Big(1- \|w\|_{W_x^{1,\frac32}} -  \|\lambda v\|_{W_x^{1,\frac32}}\Big)^{-1}\|\psi_0\|_{H_x^1} \,,
\eeqn
as claimed in \eqref{eq-psisol-unif-bd-1}.

Finally, we note that the bound \eqref{eq-Linfty-H1-psi-1} implies that    
\eqn
    |\partial_t X_\psi(t)| &=& |k-(\psi,i\nabla\psi)(t)|
    \nonumber\\
    &\leq&
    |k|+\|\psi\|_{L_t^\infty H_x^1(\R\times\R^3)}
    \nonumber\\
    &\leq& |k| + \Big(1- \|w\|_{W_x^{1,\frac32}} -  
    \|\lambda v\|_{W_x^{1,\frac32}}\Big)^{-1}\|\psi_0\|_{H_x^1} 
    \;\;,\;t\in\R \,.
\eeqn
Thus, the momentum of the tracer particle is uniformly bounded in time.
\endprf



\section{Proof of convergence to the mean field dynamics}
\label{sec-meanfield-proof-1}

\subsection{Determination of $\cL_N^{\phi_t}(k)$}
\label{ssec-cL-1}

In the following Lemma, we determine the explicit form of the operator \eqref{eq-cL-def-1}.
We will use the notation $\phi_t(x)\equiv \phi(t,x)$, similarly as in Section \ref{sec-coh-mean-lim-1}.

\begin{lemma}\label{lm-cL-1}
The selfadjoint operator $\cL_N^{\phi_t}(k)$ in \eqref{eq-cL-def-1} is given by
\eqn\label{eq-cL-expl-1}
    \cL_N^{\phi_t}(k)&=&
    \frac1{2\sqrt N}\Big(\Pb\cdot\big( a^+(i\nabla\phi_t)  +  a(i\nabla\phi_t) \big)  
    +\big( a^+(i\nabla\phi_t) +  a(i\nabla\phi_t) \big) \cdot\Pb\Big)
    \nonumber\\
    &&+ 
    \frac1{2N}\Pb^2
    \nonumber\\ 
    &&  
    + \frac{\lambda}{\sqrt N}
    \int v(x-y)  a_x^+\Big(  \overline{\phi_t(y)} a_y  
    + \phi_t(y)a^+_y  \Big) a_x \, dx dy
    \nonumber\\
    &&  
    + \frac{\lambda}{2N}
    \int v(x-y)  \, a^+_x a^+_y a_y a_x \, dx dy \,.
    \nonumber
\eeqn   
\end{lemma}

\prf
We recall from \eqref{eq-cL-def-1} that
\eqn
    \cL_N^{\phi_t}(k) 
    &=&\cW^*[\sqrt N\phi_t] \cH_N(k) \cW[\sqrt N\phi_t] - \partial_t S(t,0)
    \nonumber\\
    &&-\cW^*[\sqrt N\phi_t]\cHHar(k)\cW[\sqrt N\phi_t] -\cHcor(k)  \,.
\eeqn
The explicit expressions for the terms on the right hand side are given by
\eqn\label{eq-WcHNW-1}
    \lefteqn{
    \cW^*[\sqrt N\phi_t] \, \cH_N(k) \, \cW[\sqrt N\phi_t]
    }
    \nonumber\\
    &=&
    \frac{N}{2} \Big(k^2 - (\phi_t,i\nabla\phi_t)^2\Big)
    + \frac{N\lambda}2 \int |\phi_t(x)|^2 v(x-y) |\phi_t(y)|^2 dx dy
    \nonumber\\
    &&+
    \cW^*[\sqrt N\phi_t] \, \Big( \, -\Big(k-(\phi_t,i\nabla\phi_t)\Big)\cdot\Pb 
    + T + W_1(0) \Big) \, \cW[\sqrt N\phi_t] 
    \nonumber\\
    &&    +\frac12\Big(a^+(i\nabla\phi_t)+a(i\nabla\phi_t)\Big)^2  
    \nonumber\\
    &&+
    \frac1{2\sqrt N}\Big(\Pb\cdot\big( a^+(i\nabla\phi_t)  +  a(i\nabla\phi_t) \big)  
    +\big( a^+(i\nabla\phi_t) +  a(i\nabla\phi_t) \big) \cdot\Pb\Big)
    \nonumber\\
    &&+
    \frac1{2N}\Pb^2  
    \nonumber\\
    &&  
    +  \lambda \sqrt N
    \int v(x-y)    |\phi_t(x)|^2 \Big( a^+_y \phi_t(y) + \overline{\phi_t(y)}a_y  \Big)dx dy
    \nonumber\\
    &&  
    + \lambda
    \int v(x-y)  |\phi_t(x)|^2   a^+_y a_y \, dx dy
    \nonumber\\
    &&  
    + \lambda
    \int v(x-y)  \phi_t(x)\overline{\phi_t(y)} \,  a^+_x a_y \, dx dy
    \nonumber\\
    &&  
    + \frac{\lambda}{2}
    \int v(x-y)    \Big( \phi_t(x) \phi_t(y) a^+_x a^+_y + 
    \overline{\phi_t(x)} \overline{\phi_t(y)}a_y a_x  \Big) dx dy
    \nonumber\\
    &&  
    + \frac{\lambda}{\sqrt N}
    \int v(x-y)  a_x^+\Big(  \overline{\phi_t(y)} a_y  
    + \phi_t(y)a^+_y  \Big) a_x \, dx dy
    \nonumber\\
    &&  
    + \frac{\lambda}{2N }
    \int v(x-y)  \Big(  a^+_x a^+_y a_y a_x  \Big)dx dy \,, 
\eeqn
and
\eqn\label{eq-WcHmfW-1}
    \lefteqn{
    \cW^*[\sqrt N\phi_t]\ \, \cHHar(k) \, \cW[\sqrt N\phi_t]
    }
    \nonumber\\
    &=&  \cW^*[\sqrt N\phi_t] \, \Big( \, -\Big(k-(\phi_t,i\nabla\phi_t)\Big)\cdot\Pb 
    + T + W_1(0) \Big) \, \cW[\sqrt N\phi_t]  
    \nonumber\\
    &&
    + N\lambda \int |\phi_t(x)|^2 v(x-y) |\phi_t(y)|^2 dx dy
    \nonumber\\ 
    &&  
    +  \lambda \sqrt N
    \int v(x-y)    |\phi_t(x)|^2 \Big( a^+_y \phi_t(y) + \overline{\phi_t(y)}a_y  \Big)dx dy
    \nonumber\\
    &&  
    + \lambda
    \int v(x-y)  |\phi_t(x)|^2 \, a^+_y a_y \, dx dy \,,
\eeqn
and 
\eqn\label{eq-cHcor-def-2}
    \cHcor
    &=&
    \frac12\Big(a^+(i\nabla\phi_t)+a(i\nabla\phi_t)\Big)^2  
    \nonumber\\
    &&
    + \lambda
    \int v(x-y)  \phi_t(x)\overline{\phi_t(y)} \, a^+_x a_y \, dx dy  
    \\
    &&   
    + \frac{\lambda}{2}
    \int v(x-y)    \Big( \phi_t(x) \phi_t(y) a^+_x a^+_y + 
    \overline{\phi_t(x)} \overline{\phi_t(y)}a_y a_x  \Big) dx dy \,,
    \nonumber
\eeqn
and
\eqn 
    \partial_t S(t,0)  &=& N  \Big( \, -\frac{1}{2} k^2 + \frac{1}{2}(\phi_t,i\nabla\phi_t)^2
    \nonumber\\
    &&     + \frac\lambda2\int |\phi_t(x)|^2 v(x-y) |\phi_t(y)|^2 dx dy
    \, \Big) \,.
\eeqn
We thus obtain
\eqn\label{eq-cL-def-1}
    \cL_N^{\phi_t}(k)&=&
    \frac1{2\sqrt N}\Big(\Pb\cdot\big( a^+(i\nabla\phi_t)  +  a(i\nabla\phi_t) \big)  
    +\big( a^+(i\nabla\phi_t) +  a(i\nabla\phi_t) \big) \cdot\Pb\Big)
    \nonumber\\
    &&+
    \frac1{2N}\Pb^2
    \nonumber\\
    &&  
    + \frac{\lambda}{\sqrt N}
    \int v(x-y)  a_x^+\Big(  \overline{\phi_t(y)} a_y  
    + \phi_t(y)a^+_y  \Big) a_x \, dx dy
    \nonumber\\
    &&  
    + \frac{\lambda}{2N}
    \int v(x-y)  \,  a^+_x a^+_y a_y a_x  \, dx dy \,,
\eeqn 
as claimed in \eqref{eq-cL-expl-1}.

We note that in order to obtain \eqref{eq-WcHNW-1}, we used the following.
Introducing the abbreviated notations
\eqn
    \cW :=  \cW[\sqrt N\phi_t]
    \;\;\;,\;\;\;
    V := a^+(i\nabla\phi_t)+a(i\nabla\phi_t)
    \;\;\;,\;\;\;
    D := (\phi,i\nabla\phi) \,,
\eeqn
it is clear that 
\eqn
    \cW^* \Pb \cW = \Pb+\sqrt N V + N D \,.
\eeqn
Therefore,
\eqn
    \lefteqn{
    \frac{1}{2N}(Nk-\cW^*\Pb\cW)^2
    }
    \nonumber\\
    &=&
    \frac{1}{2N}(N(k-D)-\cW^*(\Pb-ND)\cW)^2
    \nonumber\\
    &=&
    \frac{N}{2}(k-D)^2 - (k-D)\cW^*(\Pb -ND)\cW +\frac1{2N}(\Pb+\sqrt N V)^2
    \\
    &=&
    \frac{N}{2}(k^2-D^2) - (k-D)\cW^*\Pb \cW +\frac1{2N}\Pb^2
    +\frac1{2\sqrt N}(\Pb V+V\Pb)+\frac1{2}V^2 \,.
    \nonumber
\eeqn
The terms on the last line are contained in the first five lines on the rhs of \eqref{eq-WcHNW-1}.
\endprf

\subsection{Estimates on $M(t)$}
\label{ssec-cL-2}

In this subsection, we prove the estimate \eqref{eq-cL-bd-1}.

\begin{lemma}\label{lm-cL-bd-1}
The following estimate holds,
\eqn
    \Big\| \, \cL_N^{\phi_t}(k) 
    \, \cV(t,0) \,\vac \, \Big\|_{\cF} \leq \frac{ C_0 }{\sqrt N} \,  e^{C_1 t} \,,
\eeqn
for constants $C_0$, $C_1$ depending on $\|v\|_{C^2}$ and $\|\phi_t\|_{L_t^\infty H_x^3([0,T)\times\R^3)}$.
\end{lemma}

\prf
Let
\eqn 
    \Qb \, := \, \int dk \, \langle k \rangle \, a_k^+ \, a_k
\eeqn
where $\langle k \rangle^2 := 1+k^2$.
Then, for any $\Psi\in\cF$, and $\alpha\geq1$,
\eqn 
    \||\Pb|^\alpha \, \Psi\| \, , \, \|\Nb^\alpha \, \Psi\| \, \leq \, \|\Qb^\alpha \, \Psi \| \,.
\eeqn
Thus, we obtain the following bounds on the individual terms in \eqref{eq-cL-def-1},
\eqn 
    \lefteqn{
    \Big\| \, \frac1{2\sqrt N}\Big(\Pb\cdot\big( a^+(i\nabla\phi_t)  +  a(i\nabla\phi_t) \big)  
    +\big( a^+(i\nabla\phi_t) +  a(i\nabla\phi_t) \big) \cdot\Pb\Big) \, \cV_t \, \vac \, \Big\|
    }
    \nonumber\\
    &&\hspace{5cm}\leq \,
    \frac{4\|\nabla\phi_t\|_{L^2}}{\sqrt N} \| \, \Qb^{3/2} \, \cV_t \, \vac \, \|
\eeqn
\eqn  
    \Big\| \, \frac1{2 N} \Pb^2 \, \cV_t \, \vac \, \Big\| 
    &\leq&
    \frac1{2N}\|\Qb^2 \, \cV_t \, \vac \, \|
\eeqn
\eqn 
    \lefteqn{
    \Big\| \, \frac{\lambda}{\sqrt N}\Big(
    \int v(x-y)  a_x^+\Big(  \overline{\phi_t(y)} a_y  
    + \phi_t(y)a^+_y  \Big) a_x \, dx dy\Big) \, \cV_t \, \vac \, \Big\|
    }
    \nonumber\\
    &&\hspace{4cm}\leq \,
    \frac{\lambda\|v\|_{L^\infty}\|\phi_t\|_{L^2}}{\sqrt N} \| \, \Nb^{3/2} \, \cV_t \, \vac \, \|
\eeqn
\eqn 
    \Big\| \, \frac{\lambda}{2N}
    \Big( \int v(x-y)  \, a^+_x a^+_y a_y a_x \, dx dy \Big) \, \cV_t \, \vac \, \Big\|
    \, \leq \, 
    \frac{\lambda\|v\|_{L^\infty}}{2 N} \| \, \Nb^{2} \, \cV_t \, \vac \, \|
\eeqn
Therefore, we obtain that
\eqn
    \Big\| \, \cL_N^{\phi_t}(k) 
    \, \cV(t,0) \,\vac \, \Big\|_{\cF}
    & \leq &
    \Big( \,  \frac{4\|\nabla\phi_t\|_{L^2}}{\sqrt N} \, + \, \frac1{2N}\, \Big) \, \|\Qb^2\cV_t\vac\|
    \, + \, \frac{2\lambda\|v\|_{L^\infty}}{\sqrt N}   \, \|\Nb^2\cV_t\vac\|
     \nonumber\\
    &\leq& \frac{C_0}{\sqrt N} e^{C_1 t} \,, 
\eeqn
using Lemma \ref{lm-Nb-bd-1} and Lemma \ref{lm-Pb-bd-1}.
\endprf

\begin{lemma}\label{lm-Nb-bd-1}
The following estimate holds,
\eqn
    \Big\| \, \Nb^\alpha 
    \, \cV(t,0) \,\vac \, \Big\|_{\cF} \leq e^{C_1 t}
    \;\;\;,\;\;\;
    \alpha=1,2 \,,
\eeqn
for a constants $C_1$ depending only on $\|\phi_0\|_{H^1}$.
\end{lemma}

\prf
We split $\cHcor$ (see \eqref{eq-cHcor-def-2} for the definition) into 
\eqn
    \cHcor=\cHcd+\cHcod
\eeqn
where the diagonal part $\cHcd$ commutes with the number operator,
\eqn\label{eq-cHcd-Nb-comm-1}
    [\cHcd,\Nb] \, = \, 0 \,,
\eeqn 
and where
\eqn
    \cHcod &=&
    \frac12\Big(a^+(i\nabla\phi_t)a^+(i\nabla\phi_t)+a(i\nabla\phi_t)a(i\nabla\phi_t)\Big)
    \\ 
    &&   
    + \frac{\lambda}{2}
    \int v(x-y)    \Big( \phi_t(x) \phi_t(y) a^+_x a^+_y + 
    \overline{\phi_t(x)} \overline{\phi_t(y)}a_y a_x  \Big) dx dy \,
    \nonumber\\
    &=:&\cHcodp+\cHcodm
\eeqn
is the off-diagonal part.

Defining 
\eqn
    \Nb(t) := \cV^*_t \Nb \cV_t  \,,
\eeqn
where for brevity, $\cV_t:=\cV(t,0)$, we have, for $\alpha=1,2$,
\eqn
    i\partial_t \Nb^\alpha(t) &=& \cV^*_t [\Nb^\alpha,\cHcod] \cV_t
    \nonumber\\
    &=& \alpha \cV^*_t (\cHcodp + (-1)^\alpha \cHcodm) \cV_t\,,
\eeqn
due to \eqref{eq-cHcd-Nb-comm-1}.

This implies that 
\eqn\label{eq-Nbalph-bd-1}
    \|\Nb^\alpha\cV_t\vac\|
    &=&
    \|\Nb^\alpha(t)\vac\|
    \nonumber\\
    &\leq& \|\Nb^\alpha(0)\vac\| + \int_0^t ds \, \| \, \partial_s\Nb(s) \vac \, \|
    \nonumber\\
    &\leq&
     \alpha \, \int_0^t \, ds \, \Big\| \, \cV^*_s (\cHcodps + (-1)^\alpha \cHcodms) \cV_s \vac \, \Big\| \,.
\eeqn
Writing
\eqn\label{eq-K-def-1}
    K_t(x,x'):= \frac12(i\nabla\phi_t)(x)(i\nabla\phi_t)(x') 
    + \frac\lambda2  v(x-y) \phi_t(x) \phi_t(y) \,,
\eeqn
we have
\eqn\label{eq-cHcodpm-def-1}
    \cHcodp = \int dx dx' \, K_t(x,x') \, a_x^+a_{x'}^+
    \;\;\;,\;\;\;
    \cHcodm = \int dx dx' \, \overline K_t(x,x') \, a_xa_{x'} \,,
\eeqn
and we will next prove that
\eqn\label{eq-cHcodms-bd-1}
    \Big\| \, \cV^*_s \, \cHcodms \, \cV_s\vac \, \Big\|
    &\leq&
    \| \, K_s \, \|_{L^2_{x,x'}} \| \, \Nb(s) \, \vac \, \|
\eeqn
and
\eqn\label{eq-cHcodps-bd-1}
    \Big\| \, \cV^*_s \, \cHcodps \, \cV_s\vac \, \Big\|
    &\leq&
    2 \, \| \, K_s \, \|_{L^2_{x,x'}} \| \, (1+\Nb(s)) \, \vac \, \|
    \nonumber\\
    &\leq&
    2 \, \| \, K_s \, \|_{L^2_{x,x'}} ( \, 1 \, + \, \| \, \Nb(s) \, \vac \, \| \, ) \,.
\eeqn
To prove \eqref{eq-cHcodps-bd-1}, we note that, using $K_s(x,x')=K_s(x',x)$,
\eqn 
    \lefteqn{
    \Big\| \, \cV^*_s \, \cHcodps \, \cV_s\vac \, \Big\|^2
    }
    \nonumber\\
    &=&
    \Bra \, \vac,\cV^*_s \, \cHcodms \, \cHcodps \, \cV_s\vac \, \Ket
    \nonumber\\
    &=&\int dxdx'dydy' \, \overline{K_s(x,x')} \, K_s(y,y')
    \Bra \vac \,,\, \cV_s^*\Big(2\delta(x-y)\delta(x'-y')
    \nonumber\\
    &&\hspace{2cm}
    + 4\delta(x'-y')a_y^+a_x
    + a_x^+a_y^+a_{y'}a_{x'}\Big) \, \cV_s \, \vac \, \Ket 
    \nonumber\\
    &=:&(I)+(II)+(III)
    \,.
\eeqn
We have
\eqn
    (I) \, = \, 2 \, \| K_s \|_{L^2_{x,x'}}^2
\eeqn
\eqn
    (II) &=& 4\int dx \, \Big\|\int dx' K_s(x,x')a_{x'} \, \cV_s \, \vac \, \Big\|^2
    \nonumber\\
    &\leq& 4\Big(\int dx \, \|K(x,x')\|_{L^2_{x'}}^2\Big) \, 
    \| \, a_{x'} \, \cV_s \, \vac \, \|_{L^2_{x'}}^2
    \nonumber\\
    &=& 4 \, \| K_s \|_{L^2_{x,x'}}^2 \, \| \, \Nb^{\frac12}(s) \, \vac \, \|^2
\eeqn
\eqn
    (III) &=&
    \Big|\int dxdx'dydy' \, \overline{K_s(x,x')} \, K_s(y,y') \,
    \Bra a_y a_x \, \cV_s \, \vac \,,\,  a_{y'}a_{x'} \, \cV_s \, \vac \, \Ket  \Big|
    \nonumber\\
    &\leq&
    \Big(\int dxdx'dydy' \, |K_s(x,x')|^2 \, |K_s(y,y')|^2 \Big)^{\frac12} \,
    \nonumber\\
    &&\hspace{1cm}
    \Big(\int dxdx'dydy' \,\Big\| a_y a_x \, \cV_s \, \vac \, \Big\|^2 
    \Big\|  a_{y'}a_{x'} \, \cV_s \, \vac \, \Big\|^2 \Big)^{\frac12}
    \nonumber\\
    &=&  \Big( \,\int dxdx' \, |K_s(x,x')|^2 \, \Big)  
    \\
    &&\hspace{1cm}
    \int dxdx' \,
    \Bra \, \cV_s \, \vac \,,\, \Big(a_x^+a_xa_{y}^+a_{y}-\delta(x-y)a_x^+a_y\Big) \, \cV_s \, 
    \vac \, \Ket  
    \nonumber\\
    &=&  \| K_s \|_{L^2_{x,x'}}^2 \, 
    \Big(\| \, \Nb(s) \, \vac \, \|^2
    \, - \, \| \, \Nb^{\frac12}(s) \, \vac \, \|^2\Big)
\eeqn
so that, using $\| \Nb^{\frac12}(s)  \vac \|^2\leq \|\Nb(s)\vac\|$ from Cauchy-Schwarz,
\eqn  
    \Big\| \, \cV^*_s \, \cHcodps \, \cV_s\vac \, \Big\|^2
    \, \leq \, 2 \,  \| K_s \|_{L^2_{x,x'}}^2 \Big( \, \| \, \Nb(s) \vac \, \| \, + \, 1 \, \Big)^2 \,.
\eeqn
This implies \eqref{eq-cHcodps-bd-1}.
Similarly, one arrives at \eqref{eq-cHcodms-bd-1}.

Therefore, \eqref{eq-Nbalph-bd-1} implies that
\eqn\label{eq-Nbalph-bd-2}
    1+\|\Nb^\alpha\cV_t\vac\|
    &=&
    1+\|\Nb^\alpha(t)\vac\|
    \nonumber\\
    &\leq& 1+\|\Nb^\alpha(0)\vac\| + \int_0^t ds \, \| \, \partial_s\Nb(s) \vac \, \|
    \nonumber\\
    &\leq&
    1+
     2\alpha \, \int_0^t \, ds \, \| K_s \|_{L^2_{x,x'}}^2
     \, \Big( \, 1 \, + \, \| \, \Nb(s) \, \vac \, \| \, \Big)   \,,
\eeqn
and by the Gronwall inequality,
\eqn\label{eq-Nbalph-bd-3}
    1+\|\Nb^\alpha\cV_t\vac\| 
    &\leq&  
     \exp\Big(4 \, \int_0^t \, ds \, \| K_s \|_{L^2_{x,x'}}^2 \, \Big) \,,
\eeqn
for $\alpha=1,2$. 

Finally,
\eqn
    \lefteqn{
    \| K_s \|_{L_t^\infty L^2_{x,x'}([0,T)\times\R^3\times\R^3)} 
    }
    \nonumber\\
    &\leq& 
    \frac12\|\nabla\phi_s\|_{L_x^2}^2+
    \frac{\lambda}2\Big(\int |\phi_s(x)|^2 \, |v(x-y)|^2 \, |\phi_s(y)|^2 dx \, dy\Big)^{\frac12}
    \nonumber\\
    &\leq&(1+\lambda\|v\|_{L_x^\infty}) \, \|\phi_t\|_{L_x^\infty H^1_{x}([0,T)\times\R^3)}^2 \,.
\eeqn
This implies the claim of the lemma.
\endprf

\begin{lemma}\label{lm-Pb-bd-1}
Let
\eqn 
    \Qb \, = \, \int dk \, \langle k \rangle \, a_k^+ \, a_k
\eeqn
where $\langle k \rangle^2 = 1+k^2$.
Then, the following estimate holds, 
\eqn
    \Big\| \, \Qb^\alpha 
    \, \cV(t,0) \,\vac \, \Big\|_{\cF} \, \leq \,  e^{C_1 t}
    \;\;\;,\;\;\;
    \alpha=1,2 \,,
\eeqn
for a constant $C_1$ depending only on $\|v\|_{C^2}$ and  $\|\phi\|_{L^\infty_t H^3([0,T)\times\R^3)}$.
\end{lemma}

\prf
We define 
\eqn
    \Qb(t) := \cV^*_t \Qb \cV_t  \,,
\eeqn
where $\cV_t:=\cV(t,0)$ as before. For $\alpha=1,2$, we find
\eqn
    i\partial_t \Qb^\alpha(t) &=& \cV^*_t [\Qb^\alpha,\cHcd+\cHcod] \cV_t \,.
\eeqn
Similarly to \eqref{eq-K-def-1} and \eqref{eq-cHcodpm-def-1}, we define
\eqn
    J_t(x,x'):= \overline{(i\nabla\phi_t)(x)}(i\nabla\phi_t)(x')  
    + \lambda v(x-y) \overline{\phi_t(x)} \phi_t(y) \,,
\eeqn
so that
\eqn 
    \cHcd = \frac12\|\nabla\phi\|_{L_x^2}^2 \, + \, \int dx dx' \, J_t(x,x') \, a_x^+a_{x'} \,.
\eeqn
Since
\eqn\label{eq-QbcH-comm-1}
    \lefteqn{
    [\Qb,\cHcd+\cHcod] \, = \, \int dx dx' a_x^* a_{x'} 
    (\langle\nabla_x\rangle-\langle\nabla_{x'}\rangle)  J_t(x,x') 
    }
    \\
    &&+ \int dx dx' \Big\{ a_x^* a^*_{x'} 
    (\langle\nabla_x\rangle+\langle\nabla_{x'}\rangle) K_t(x,x') \, - \, 
    a_x a_{x'}  (\nabla_x+\nabla_{x'}) \overline{ K_t(x,x') } \Big\}
    \nonumber
\eeqn
Then, the estimate
\eqn\label{eq-partt0-Qb-bd-1}
    \lefteqn{
    \partial_t\|\Qb(t)\vac\| 
    }
    \nonumber\\
    &\leq& C \, 
    \Big(\, \|(\langle\nabla_x\rangle-\langle\nabla_{x'}\rangle) J_t\|_{L^2_{x,x'}} 
    \nonumber\\
    &&
    \hspace{1.5cm}
    + \|(\langle\nabla_x\rangle+\langle\nabla_{x'}\rangle) K_t\|_{L^2_{x,x'}}
    \Big) \, (\|\Nb(t)\vac\|+1)
    \\
    &\leq& C \, \Big(  \|(\langle\nabla_x\rangle-\langle\nabla_{x'}\rangle) J_t\|_{L^2_{x,x'}}  
    + \|(\langle\nabla_x\rangle+\langle\nabla_{x'}\rangle) K_t\|_{L^2_{x,x'}}
    \Big)  \, (\|\Qb(t)\vac\|+1)
    \nonumber
\eeqn
follows from the same arguments as the proof of \eqref{eq-cHcodms-bd-1}, \eqref{eq-cHcodps-bd-1}.

For $\alpha=2$, we have
\eqn 
    \partial_t\|\Qb^2\cV_t\vac\|
    &\leq& \|[\Qb^2,\cHcd+\cHcod]\cV_t\vac\| \,.
\eeqn
Taking the commutator with an operator acts as a derivation, thus  
\eqn 
    \lefteqn{
    [\Qb^2,\cHcd+\cHcod] 
    }
    \nonumber\\
    &=&
    \Qb \, [\Qb \, , \, \cHcd+\cHcod] \, + \, [\Qb \, , \, \cHcd+\cHcod] \, \Qb
    \nonumber\\
    &=&
    2 [\Qb \, , \, \cHcd+\cHcod] \Qb \, + \, [\Qb \, , \, [\Qb \, , \, \cHcd+\cHcod]]
\eeqn
where, similarly to \eqref{eq-QbcH-comm-1}, 
\eqn 
    \lefteqn{
    [\Qb \,,\, [\Qb,\cHcd+\cHcod]]   
    }
    \nonumber\\
    &=&
    \int dx dx' a_x^* a_{x'} (\langle\nabla_x\rangle-\langle\nabla_{x'}\rangle)^2 J_t(x,x') 
    \\
    &&+ \int dx dx' \Big\{ a_x^* a^*_{x'}  
    (\langle\nabla_x\rangle+\langle\nabla_{x'}\rangle)^2 K_t(x,x') \, 
    \nonumber\\
    &&\hspace{4cm}
    + \, 
    a_x a_{x'}  (\langle\nabla_x\rangle+\langle\nabla_{x'}\rangle)^2 \overline{ K_t(x,x') } \Big\} \,.
    \nonumber
\eeqn
In analogy to \eqref{eq-partt0-Qb-bd-1}, we therefore find that
\eqn 
    \partial_t\|\Qb^2\cV_t\vac\| &\leq&A_1(t)(\| \Qb^2 \cV_t \vac\|+1)+A_2(t)(\| \Qb \cV_t \vac\|+1)
\eeqn
where
\eqn 
    A_1(t) &:=& C \, \Big(  \|(\langle\nabla_x\rangle-\langle\nabla_{x'}\rangle) J_t\|_{L^2_{x,x'}}  
    + \|(\langle\nabla_x\rangle+\langle\nabla_{x'}\rangle) K_t\|_{L^2_{x,x'}}
    \Big)  
    \nonumber
\eeqn
and 
\eqn 
    A_2(t) &:=& 
    C \, \Big(  \|(\langle\nabla_x\rangle-\langle\nabla_{x'}\rangle)^2 J_t\|_{L^2_{x,x'}}  
    + \|(\langle\nabla_x\rangle+\langle\nabla_{x'}\rangle)^2 K_t\|_{L^2_{x,x'}}
    \Big)  \,.
    \nonumber
\eeqn
Since $\|\Qb\cV_t\vac\|\leq\|\Qb^2\cV_t\vac\|$, we conclude that
\eqn 
    \partial_t\|\Qb^2\cV_t\vac\| &\leq&(A_1(t)+A_2(t))(\| \Qb^2 \cV_t \vac\|+1) 
\eeqn
and hence,
\eqn 
    \|\Qb^2\cV_t\vac\|
    &\leq&
    \exp\Big( \, \int_0^t ds (A_1(s)+A_2(s))\, \Big) 
    \nonumber\\
    &\leq&
    \exp\Big( \, t \, \|A_1+A_2\|_{L_t^\infty([0,T))} \, \Big) \,,
\eeqn
using that $\|\Qb^2\cV_0\vac\|=0$, due to $\cV_0=\1$.

Finally, we have, for $\alpha=1,2$,
\eqn 
    \lefteqn{
    \|(\langle\nabla_x\rangle-\langle\nabla_{x'}\rangle)^\alpha J_t\|_{L_t^\infty L^2_{x,x'}
    ([0,T)\times\R^3\times\R^3)}
    }
    \nonumber\\
    &&\hspace{1cm}\, + \,
    \|(\langle\nabla_x\rangle+\langle\nabla_{x'}\rangle)^\alpha K_t\|_{L_t^\infty L^2_{x,x'}
    ([0,T)\times\R^3\times\R^3)}
    \nonumber\\
    &&\hspace{3cm}
    \, \leq \, (1+\lambda\|v\|_{C^\alpha(\R^3)})\|\phi_t\|_{L_t^\infty H_x^{1+\alpha}([0,T)\times\R^3)}^2 \,, \;\;
\eeqn
as can be easily checked.
\endprf

\noindent
{\bf Acknowledgements:} 
Part of this work was done while A. Soffer was a visiting professor at CCNU (Central China Normal University). 
A. Soffer is partially supported by NSFC grant No.11671163 and NSF grant DMS01600749. 
Part of this work was done when T. Chen was visiting CCNU.
T. Chen thanks A. Soffer for his hospitality at CCNU in Wuhan.
T. Chen gratefully acknlowledges support by the NSF through grants DMS-1151414 (CAREER) and DMS-1716198.

\end{document}